\documentclass[%
 aip,
 amsmath,amssymb,
 reprint,%
]{revtex4-1}

\usepackage{graphicx}
\usepackage{dcolumn}
\usepackage{bm}

\usepackage[normalem]{ulem}
\usepackage{siunitx}
\usepackage{physics}
\usepackage{pgfplots}
\usepackage{soul}
\pgfplotsset{compat=1.18}
\usepackage{romannum}
\usepackage[utf8]{inputenc}
\usepackage[T1]{fontenc}
\usepackage{mathptmx}
\usepackage{etoolbox}

\makeatletter
\def\@email#1#2{%
 \endgroup
 \patchcmd{\titleblock@produce}
  {\frontmatter@RRAPformat}
  {\frontmatter@RRAPformat{\produce@RRAP{*#1\href{mailto:#2}{#2}}}\frontmatter@RRAPformat}
  {}{}
}%
\makeatother
\begin{document}
\preprint{AIP/123-QED}

\title{Bending elasticity of the reversible freely-jointed chain}
\author{Minsu Yi}
\author{Dongju Lee}
\author{Panayotis Benetatos}
\email{alstn8588@knu.ac.kr, pben@knu.ac.kr}
\affiliation{ 
Department of Physics, Kyungpook National University
}

\date{\today}

\begin{abstract}
The freely jointed chain model with reversible hinges (rFJC) is the simplest theoretical model which captures reversible transitions of the local bending stiffness along the polymer chain backbone ({\it e.g.}, helix-coil-type of local conformational changes or changes due to the binding/unbinding of ligands). In this work, we analyze the bending fluctuations and the bending response of a grafted rFJC in the Gibbs (fixed-force) ensemble. We obtain a recursion relation for the partition function of the grafted rFJC under bending force, which allows, in principle, exact-numerical calculation of the behavior of a rFJC of arbitrary size. In contrast to stretching, we show that under sufficiently stiff conditions, the differential bending compliance and the mean fraction of closed hinges are non-monotonic functions of the force. We also obtain the persistence length $L_p$ of the rFJC, the moments \(\langle R^2 \rangle\) (mean-square end-to-end distance) and \(\langle z^2 \rangle\) (mean-square transverse deflection) for the discrete chain and take the continuum limit. The tangent vector auto-correlation decays exponentially, as in the wormlike chain model (WLC). Remarkably, the expression of \(\langle R^2 \rangle\) as a function of the contour length $L$  becomes the same as that in the WLC. In the thermodynamic limit, we have calculated the exact bending response analytically. As expected,  for $L\gg L_p$ the boundary conditions do not matter, and the bending becomes equivalent to stretching. In contrast, for $L_p\gg L$, we have shown the non-monotonicity of the bending response (the compliance and mean fraction of closed hinges).
\end{abstract}

\maketitle

\section{\label{sec:Introduction}Introduction}

Various biopolymers, including DNA and polypeptides, besides their backbone conformations, exhibit internal degrees of freedom. This affects their macroscopic behavior, and its analysis involves interesting statistical mechanics. Some of the internal degrees of freedom are associated with the local bending stiffness, and this affects the bending stiffness of the whole system, which determines the bending elasticity.

Many important biopolymers, such as dsDNA and F-actin, are semiflexible and their behavior is dominated by their bending stiffness~\cite{Sung_book,Schiessel_book}. This results in a persistence of the orientation along the polymer, quantified by the relevant correlation length, known as the persistence length \(L_p\). For dsDNA, $L_p \approx  \SI{47}{\nano\metre}$, while for F-actin, $L_p \approx  \SI{17}{\micro\metre}$.\cite{gittes1993flexural, wang1997stretching} There are various theoretical models of semiflexible polymers. One of the most widely used is the wormlike chain (WLC), a locally inextensible, one-dimensional curve with bending stiffness. It is the continuum version of the Kratky-Porod chain (KP chain).\cite{KP_original} The KP model assigns a harmonic bending potential to each bond, where the aligning is energetically favored. The simplest model is the freely jointed chain (FJC), which consists of a sequence of rigid rods linked together by freely rotating hinges~\cite{Rubinstein_polymer}. In this model, the rod segments have infinite bending stiffness, whereas the hinges have zero bending stiffness. Even though the stretching response of double stranded DNA better agrees with the predictions of the WLC model~\cite{Marko-Siggia}, stretching single stranded DNA is better described by the FJC because strong stretching is dominated by the short wavelength undulations that reveal the discrete nature of the bonds~\cite{Dobrynin_DNA_FJC,Saleh_DNA_FJC}.  

Apart from the fluctuating bending (curvature of the backbone), there may be other internal degrees of freedom, associated with a reversible bending stiffness. The reversible change in the local bending stiffness may be due to internal conformational changes or due to the interaction with the environment. An example of the former is the helix-coil transition in polypeptides,~\cite{Grosberg} and an example of the latter is the reversible binding/unbinding of DNA-binding proteins along the backbone of single or double stranded DNA.~\cite{DNA_Protein_Bruinsma,DNA_Protein_experiments,Andelman_PRE,Yan_Marko_PRE,Zhang_Marko_PRE} At a coarse-grained level, these reversible changes can be viewed as fluctuations between two different values of the local bending stiffness.

Several theoretical studies have dealt with the elasticity of polymers with locally fluctuating bending stiffness. In Refs.~\onlinecite{Buhot_PRL,Tamashiro_Pincus_PRE}, the chain is modeled as a FJC with rods of different length, depending on whether an elementary rod belongs to a helix or a coil block of the polypeptide. The rod length becomes an internal degree of freedom. In Refs.~\onlinecite{chakrabarti2005nonlinear,Chakrabarti_Levine_simulation}, a discrete WLC model is used, and the bending stiffness of each bond is a fluctuating degree of freedom. Ref.~\onlinecite{StretchingrFJC_gibbs} presents one of the simplest models, the reversible freely jointed chain (rFJC), and its stretching elasticity in the Gibbs ensemble is analyzed. It is a freely jointed chain with reversible hinges. An open hinge is the same as in the usual FJC, whereas a closed hinge links two neighboring rods and forms a rod with twice the original length. In fact, the same model was first introduced and analyzed in the Helmholtz ensemble by Winkler {\it et al.}~\cite{Winkler_FJC,Winkler_FJC_macromolecules} 

The presence of reversible hinges not only alters the tensile elasticity of the FJC, but also introduces orientational correlations and bending stiffness along distances longer than the length of the elementary rod. As such, it becomes a bona fide semiflexible chain. The bending elasticity of the rFJC is the subject of the present article. We investigate both the response to a bending force and the bending conformations. A quantity of central interest is the emergent persistence length.

This article is organized as follows. In Section II, we introduce our model. In Section III, we present some exact results for the simple cases of an rFJC with one or two reversible hinges. In Section IV, we derive a recursion relation for the Gibbs partition function of an rFJC with N reversible hinges, and use it to calculate the bending elasticity for a chain with 13 reversible hinges. In Section V, we derive and present a mean field approximation for the partition function, which significantly facilitates our calculations. In Section VI, we derive and analyze the continuum limit of the rFJC. We also discuss the rod-like limit. In Section VII, we use the generating function method (necklace model) to obtain the exact force-deflection relation in the thermodynamic limit. We summarize in Section VIII.

\section{\label{sec:Model}Model}
\begin{figure}
    \includegraphics[width=0.48\textwidth]{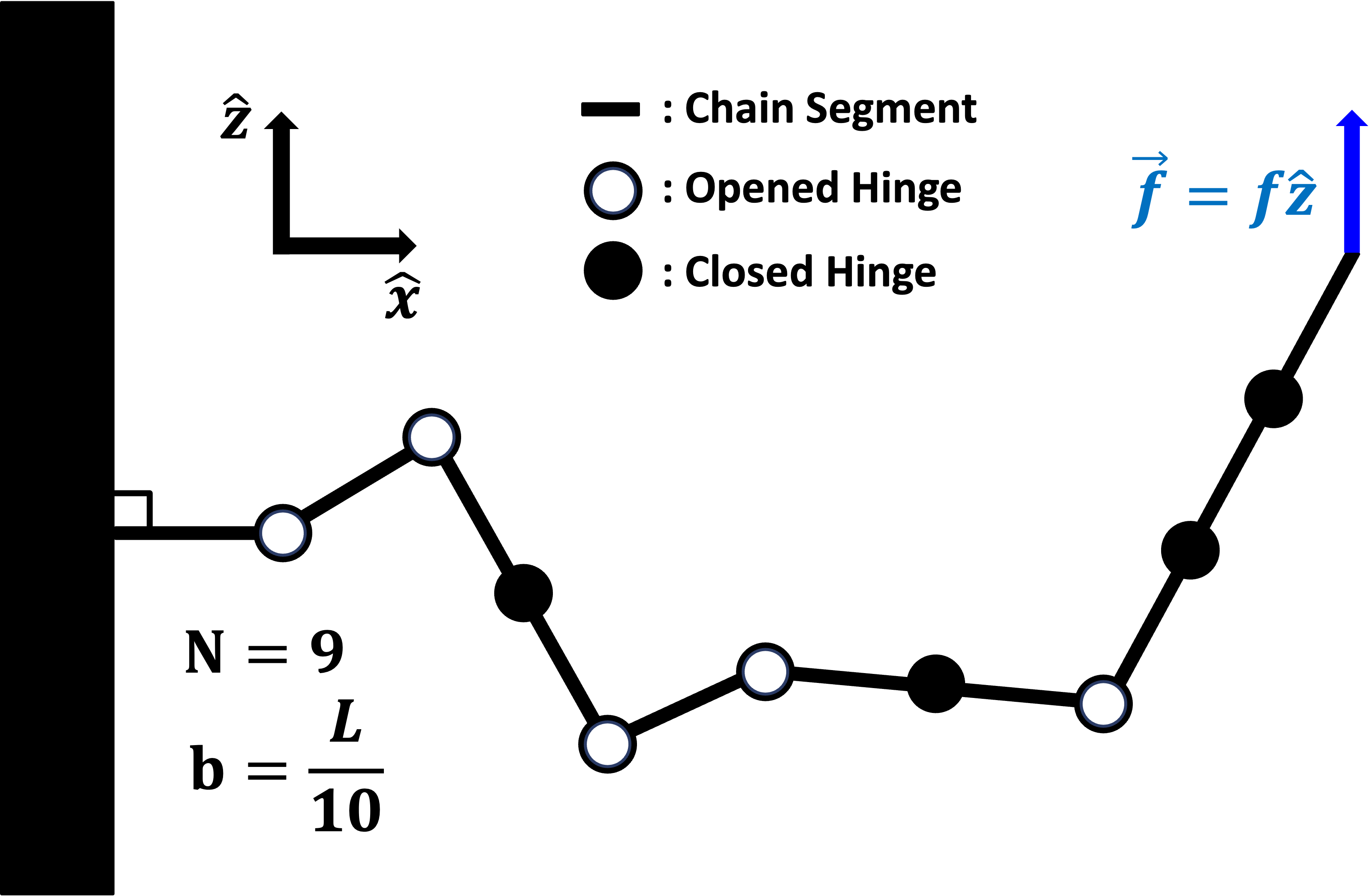}
    \caption{\label{fig:intro_bending_rFJC} A schematic illustration of the grafted rFJC. In this microstate, four hinges are closed, and five hinges are opened. Notice that we consider the chain to be grafted, and we apply a bending force at the other end. The wall simply provides rigid support to the grafted end and does not exert any steric repulsion.}
\end{figure}

Our model consists of \(N+1\) chain segments which are each one-dimensional massless rigid rods of length \(b\) linked together with \(N\) reversible hinges. We consider the chain to be grafted perpendicular to a wall and investigate its bending elasticity. One end of the chain is fixed to the "wall" in the \(\hat{x}\) direction, and the first segment cannot have any fluctuations. Moreover, the chain is considered in three dimensions, which allows any orientation of the segments along the unit sphere. The "wall" simply provides a grafting point and does not impose any other constraint (it is not impenetrable). 

A constant bending force \(f\) is applied at the other end of the chain, perpendicular to the grafting direction. This means that we are treating the bending elasticity in the Gibbs ensemble (the force is the fixed control parameter, and the extension of the chain can fluctuate). A description of the model is shown in FIG.~\ref{fig:intro_bending_rFJC}. In this paper, as in Ref.~\onlinecite{StretchingrFJC_gibbs}, the abbreviation rFJC refers to the Freely Jointed Chain with reversible hinges, and uFJC refers to the usual Freely Jointed Chain with permanently opened hinges.

We defined \(N\) as the number of reversible hinges, so the total contour length of the chain is \(L = (N+1)b\), including the first grafted segment. An activation energy parameter \(\epsilon\)  controls the opening-closing of the reversible hinges. When a hinge is opened, it allows the two connected segments to have any relative orientation, whereas when it is closed, they get aligned and act as a single rigid rod. This mechanism on hinges captures the reversible transitions of the local bending stiffness.

Each microstate is characterized by the orientation \(\Omega_i = \{\theta_i,\phi_i\}\) of each chain segment $i$ for all \(i = 1,\,2,\, \dots\,,\, N+1\) and the opened or closed state of each hinge which is parameterized by the occupation number \(n_i = 0\) or \(1\), respectively,  where \(i =  1,\,2,\, \dots\,,\, N\). The angles (polar, $\theta_i$, and azimuthal, $\phi_i$) are measured with respect to the $\hat{ z}$ direction. Setting \(\vec{f} = f\hat{z}\), the hamiltonian of the system is given by
\begin{eqnarray}
    H = -\sum_{i=2}^{N+1} fb\cos\theta_i -\sum_{i=1}^{N} n_i\epsilon
    \label{eq:hamiltonian}.
\end{eqnarray}
Notice that we have ignored the contribution of the kinetic energy. The details are discussed in Appendix.~\ref{sec:kinetic_appendix}. The hamiltonian yields the partition function
\begin{eqnarray}
    Z(N) = \prod_{i=2}^{N+1}\sum_{n_{i-1}=0}^{1}\int\,d\Omega_i \exp\Biggr(\frac{fb \cos\theta_i}{k_BT}\Biggr)\nu(n_{i-1})
    \label{eq:partit_func},
\end{eqnarray}
where
\begin{eqnarray}
    \nu(n_i) &=&  \ \exp\Biggr(\frac{n_i\epsilon}{k_BT}\Biggr)\frac{\delta(\theta_{i+1}-\theta_{i})\delta(\phi_{i+1}-\phi_{i})}{\sin\theta_i} \ \text{(}n_i=1\text{)} \qquad \nonumber \\ \nonumber\\
    &=& 1 \ \qquad \qquad \qquad \qquad \qquad \qquad \qquad \; \, \, \, \; \text{(}n_i=0\text{)}
    \label{eq:nu}
\end{eqnarray}
and \(\theta_1=\pi/2\). Note that if \(n_i=1\)\, we have the Boltzmann weight of the closed hinge and a delta function, where the delta function makes the angle of the two adjacent segments to be the same. Also, since the first segment is grafted, the product of Eq.~(\ref{eq:partit_func}) runs from \(i=2\) to \(i=N+1\), and \(\theta_1\) is always fixed as \(\pi/2\) (perpendicular to the wall).

The energy parameter $\epsilon$ also appears in models of other bistable systems. A more detailed discussion about its meaning can be found in Ref.~\onlinecite{gnoh_flexible}.

It can be trivially shown that if all the hinges are opened (\(n_i = 0\) for all \(i\) or \(\epsilon \rightarrow -\infty\)), we always recover the partition function for the grafted usual Freely Jointed Chain (uFJC).
\begin{eqnarray}
    Z_{uFJC}=\Biggr[\frac{4\pi\sinh(fb/k_BT)}{fb/k_BT}\Biggr]^N
    \label{eq:partit_func_ufjc}.
\end{eqnarray}

Lastly, we mention that fixing the occupation number of the first hinge as \(n_1=0\) in Eq.~(\ref{eq:partit_func}), it reduces to the partition function of an rFJC under stretching with \(N-1\) hinges in Ref.~\onlinecite{StretchingrFJC_gibbs}.

\section{\label{sec:Exact_Results}Exact Results}

In this Section, we consider a small system containing a few hinges and calculate the bending elasticity exactly. The following results exhibit interesting properties, and we expect that some of these properties could persist for large \(N\) without losing generality.

\subsection{\label{sec:One_Rev}One Reversible Hinge}

We start our analysis from the simplest case. Let us consider a chain with one reversible hinge. In this case, the partition function is

\begin{eqnarray}
    Z(1) = e^{\epsilon}+\frac{4\pi\sinh(f)}{f}
    \label{eq:partit_one}.
\end{eqnarray}
For the sake of simplicity, we set \(b = 1\) and \(k_BT = 1\). If we let \(\epsilon \rightarrow -\infty\), we recover the partition function of uFJC as we expected from Eq.~(\ref{eq:partit_func_ufjc}).

The mean occupation number (or the mean fraction of closed hinges in the statistical ensemble) is given by
\begin{eqnarray}
    \langle n \rangle_1 &=&\frac{\partial\ln Z(1)}{\partial \epsilon}\nonumber
    \\&=&e^{\epsilon}\Biggr/\Biggr[e^{\epsilon}+\frac{4\pi\sinh(f)}{f}\Biggr]
    \label{eq:mean_occ_one}.
\end{eqnarray}

When \(f = 0\), the mean occupation number is simply \(\langle n \rangle_1 = e^{\epsilon}/(e^{\epsilon}+4\pi) \). In the denominator, we have the contribution from the closed hinge, which is \(e^{\epsilon}\), and \(4\pi\) from the opened hinge due to all possible orientations in three dimensions. This geometric degree of freedom underpins the difference from the simple Langmuir adsorption model.\cite{langmuir1918isotherm} Notice also that if \(f\rightarrow \infty\), then \(\langle n \rangle_1\rightarrow 0\). Since the first segment is grafted, the large force always opens the first hinge.

The force-deflection relation is given by
\begin{eqnarray}
    \langle z \rangle_1 &=&\frac{\partial\ln Z(1)}{\partial f}\nonumber
    \\&=&\frac{4\pi\cosh(f)/f-4\pi\sinh(f)/f^2}{e^{\epsilon}+4\pi\sinh(f)/f}
    \label{eq:mean_ext_one}.
\end{eqnarray}
In the case of stretching, the mean extension of the rFJC at a certain force is greater than that of the the uFJC. Increasing the elongational force, effectively stiffens the chain, and it tends to behave more as a length of \(2b\) rigid rod.\cite{StretchingrFJC_gibbs} However, the rFJC can become less compliant in the bending case than the uFJC because the activation energy resists opening a hinge due the bending force.

To examine the bending response, we define the differential bending compliance for \(N\) hinges as
\begin{eqnarray}
    C_N := \frac{1}{Nb}
    \frac{\partial \langle z \rangle_N}{\partial f}
    \label{eq:comp}.
\end{eqnarray}
In the small force regime \(f \ll k_BT/(Nb)\), the compliance of the rFJC becomes
\begin{eqnarray}
    C_1 = \frac{4\pi}{3 (e^\epsilon + 4\pi)}
    \label{eq:comp_rfjc_one_small}.
\end{eqnarray}
The compliance above is always smaller than the value of \(1/3\), where the value \(1/3\) corresponds to the linear compliance of the uFJC~\cite{Rubinstein_polymer}. In the large force regime (\(f \gg k_BT/b\), \(fb \gg \epsilon\)), we have
\begin{eqnarray}
    \frac{\langle z \rangle_1}{b} = 1-\frac{k_BT}{fb}
    \label{eq:comp_rfjc_one_large}.
\end{eqnarray}
This expression is the same as the compliance obtained for stretching a rFJC with a single segment. Since \(\epsilon\) is small and finite compared to \(fb\), the large force always opens the hinge.

\subsection{\label{sec:Two_Rev}Two Reversible Hinges}

The partition function for two reversible hinges is
\begin{eqnarray}
    Z(2) = e^{2\epsilon}&+&e^{\epsilon}\frac{4\pi\sinh(f)}{f} \ + \ \Biggr(\frac{4\pi\sinh(f)}{f}\Biggr)^2\nonumber
    \\ &+&e^{\epsilon}\frac{4\pi\sinh(2f)}{2f}
    \label{eq:partit_two}.
\end{eqnarray}
It is reassuring that by taking the limit \(\epsilon \rightarrow -\infty\), we recover the partition function of uFJC. Also, assuming the first hinge is always opened (\(n_1 = 0\)), we obtain the partition function for stretching the \(N = 1\) rFJC. This refers to \((4\pi \sinh(f)/f)^2+e^{\epsilon} 4\pi \sinh(2f)/2f\) in the RHS. The partition function above yields the mean occupation number and the mean deflection as below:
\begin{widetext}
    \begin{eqnarray}
    \langle n \rangle_2
    &=&\frac{1}{2}\frac{\partial\ln Z(2)}{\partial \epsilon}
    =\frac{1}{2}
    \frac{2e^{2\epsilon}+e^{\epsilon}4\pi\sinh(f)/f+e^{\epsilon}4\pi\sinh(2f)/2f}{e^{2\epsilon}+e^\epsilon4\pi\sinh(f)/f+(4\pi\sinh(f)/f)^2+e^\epsilon4\pi\sinh(2f)/2f} \nonumber \\ \nonumber \\
    \langle z \rangle_2&=&\frac{e^{\epsilon}4\pi(\cosh(f)/f-\sinh(f)/f^2)+(4\pi)^2(2\cosh(f)\sinh(f)/f^2-2\sinh^2(f)/f^3)+e^{\epsilon} 4\pi(\cosh(2f)/f-\sinh(2f)/2f^2)}
    {e^{2\epsilon}+e^\epsilon4\pi\sinh(f)/f+(4\pi\sinh(f)/f)^2+e^{\epsilon}4\pi\sinh(2f)/2f} \nonumber \\
    \label{eq:mean_two}
    \end{eqnarray}
\end{widetext}

In the strong bending regime, the microstate with \(n_1=0\) and \(n_2=1\) whose Boltzmann weight is \(e^\epsilon4\pi\sinh(2f)/2f\) dominates. In other words, the last two segments stiffen and form a length \(2b\) rod. Therefore, we easily understand \(\langle n \rangle \rightarrow 1/2\) in this regime.

\begin{figure}
    \includegraphics[width=0.48\textwidth]{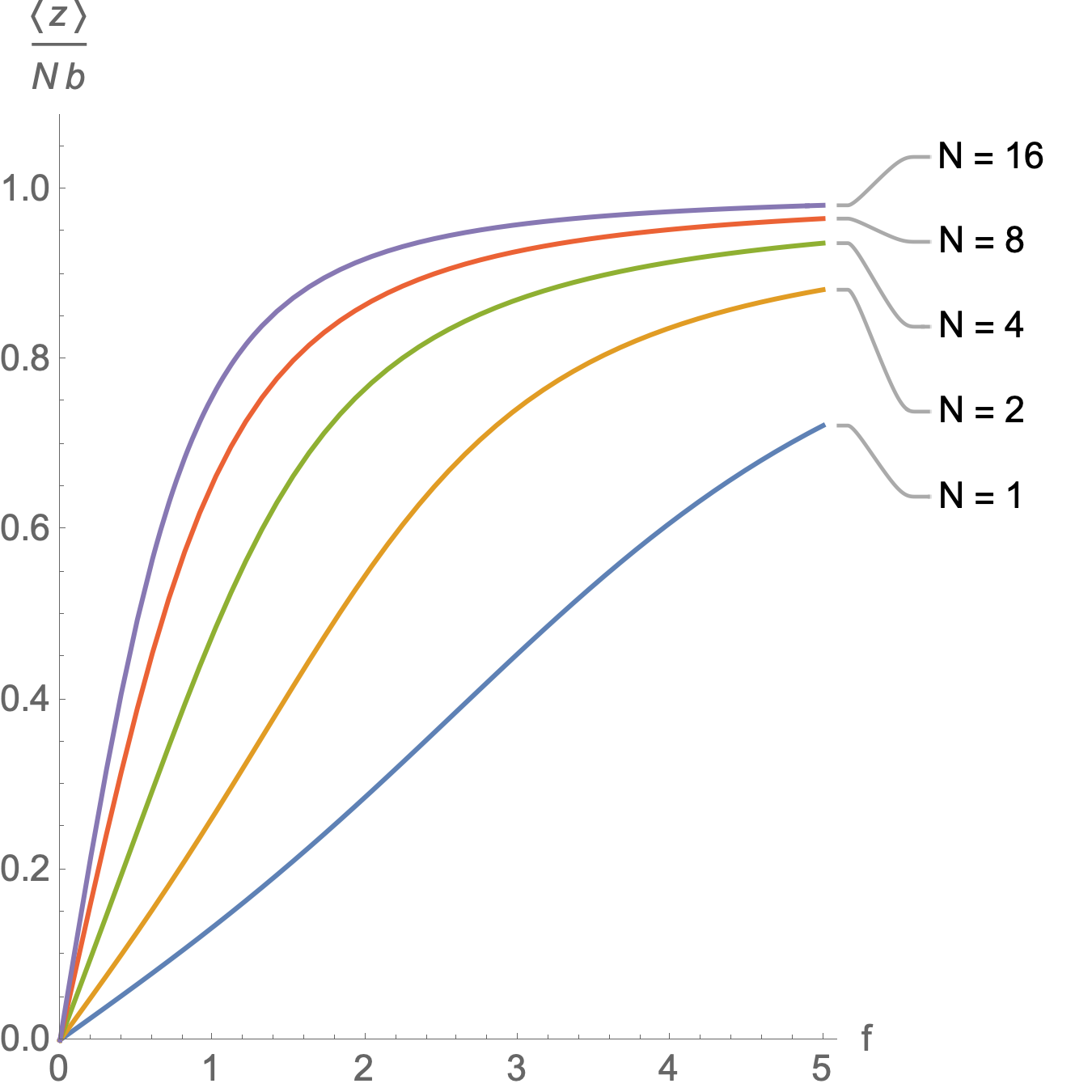}
    \caption{\label{fig:comp_one_two_small} Mean deflection \(\langle z\rangle\) divided by \(Nb\), for various values of the number of hinges, as a function of the bending force \(f\). Since the lever-arm effect takes part, the rFJC with more hinges is more easily bent. \(\epsilon = 3\), \(k_BT = 1\), \(b = 1\).}
\end{figure}

Now consider the force-deflection relation. We mentioned in Section~\ref{sec:One_Rev} that the existence of the reversible hinge makes the chain resist being bending. However, the situation changes when \(N \geq 2\). In the case of \(N \geq 2\), the lever-arm effect starts to play an important role. When \(N = 2\), for example, the existence of the hinge provides the possibility to form a longer rod of length \(2b\), {\it i.e.} the microstate with \(n_1 = 0\), \(n_2 = 1\). The contribution to the partition function of each microstate is
\begin{eqnarray}
    Z(2)(n_1=1,n_2=1)&=&Z(2)(1,1) \;=\; e^{2\epsilon}, \qquad \nonumber
    \\ Z(2)(1,0)&=&e^{\epsilon}\frac{4\pi\sinh(f)}{f}, \nonumber
    \\ Z(2)(0,0)&=&\Biggr(\frac{4\pi\sinh(f)}{f}\Biggr)^2,\nonumber
    \\ Z(2)(0,1)&=&e^{\epsilon}\frac{4\pi\sinh(2f)}{2f} 
    \label{eq:leverarm}.
\end{eqnarray}
Notice that \(Z(2)(0,1) = \mathcal{O}(e^f)\times Z(2)(1,0)\) and \(Z(2)(0,1) = \mathcal{O}(f)\times Z(2)(0,0)\). If \(f\) increases, keeping \(\epsilon\) finite, \(Z(2)(0,1)\) grows the fastest among Eq.~(\ref{eq:leverarm}). Now, we compare \(N = 1\) and \(N = 2\) to verify the lever-arm effect. Recall from Eq.~(\ref{eq:mean_ext_one}) that the existence of the hinge provides the possibility to form a longer rod, so it makes the rFJC less compliant when \(N=1\). In that case, the contribution to the partition function by opening and closing of the hinge are each \(4\pi \sinh(f)/f\) and \(e^\epsilon\). However, we have an exponentially larger contribution for \(N = 2\) due to the bending force when the hinge is closed. This exponentially larger contribution refers to the term \(Z(2)(0,1)=e^\epsilon4\pi \sinh(2f)/2f\). Thus, if \(N\) increases, the bent microstates sample exponentially larger phase space by \(\mathcal{O}(e^{Nf})\) in the partition function, and this is the lever-arm effect in the rFJC. This is graphically shown in FIG.~\ref{fig:comp_one_two_small}.

\begin{figure}
    \includegraphics[width=0.48\textwidth]{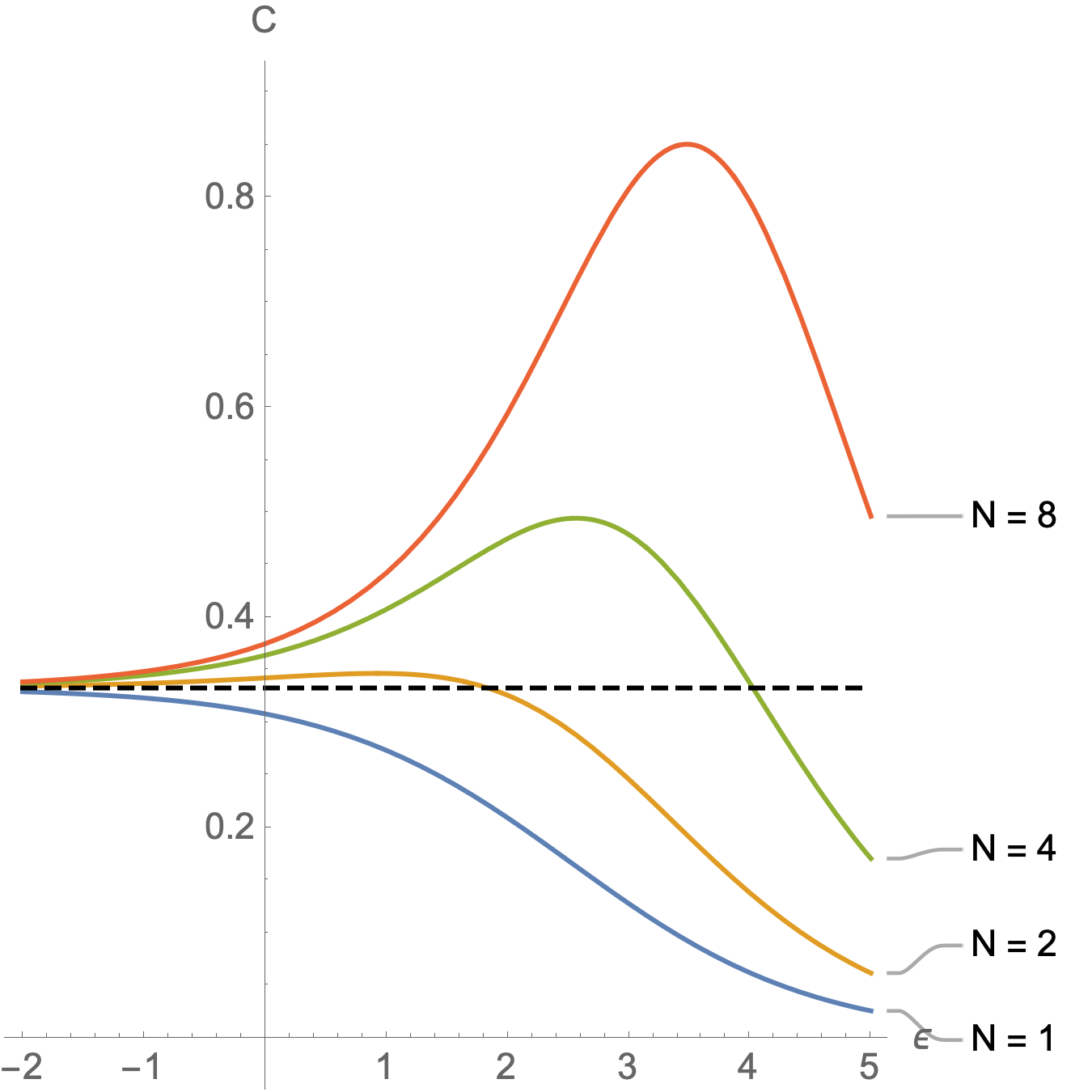}
    \caption{\label{fig:comp_n} The differential compliance at \(f=0\) (linear response), for various values of the number of hinges, as a function of \(\epsilon\). Since the lever-arm effect gets involved, the rFJC with more hinges is more easily bent for any \(\epsilon\), provided that $\epsilon$ is not very large . Notice that it converges to \(1/3\) (the dashed line) when \(\epsilon \rightarrow -\infty\). \(k_BT = 1\), \(b = 1\).}
\end{figure}

To show the lever-arm effect quantitatively, we have considered the bending compliance for \(N = 2\). The bending compliance for \(N = 2\) in the small force regime \(f \ll k_BT/(2b)\) is given by
\begin{eqnarray}
C_2=\frac{4\pi}{3(e^\epsilon+4\pi)}+\frac{4\pi e^\epsilon}{2(e^\epsilon+4\pi)^2}
\label{eq:comp_rjfc_two_small}.
\end{eqnarray}
Since the chain for $N=2$ can have a longer rod that bends, we get an extra term compared to \(N = 1\), Eq.~(\ref{eq:comp_rfjc_one_small}). Our numerical investigation in FIG.~\ref{fig:comp_one_two_small} and FIG.~\ref{fig:comp_n} shows that, as long as $\epsilon$ is not very large, the lever-arm effect persists for any value of $N>1$. As we discuss later in the article, $\epsilon$ gives rise to an effective persistence length of the rFJC. When that persistence length is larger than the total contour length of the chain, $Nb$, the linear bending compliance is less than that of the uFJC. If, on the other hand, the value of $\epsilon$ is such that the persistence length is less than the total contour length, we will always get a lever-arm effect. What happens when $N\gg1$? In Section~\ref{sec:Generating_Function}, we calculate the linear response in the thermodynamic limit, Eq.~(\ref{eq:approx_def_small}), and we get a linear compliance greater than that of the uFJC (1/3). Even though the net effect is mathematically the same, in the thermodynamic limit, we would not call it "lever-arm" effect. In that case, bending and stretching become equivalent. Linear response theory requires linear compliance to be proportional to \(\langle z^2 \rangle\), which is larger than that of a FJC with the same contour length and larger Kuhn length. So, irrespective of the value of $N$, we get this increase in the linear compliance.
Besides, it is interesting to point out that the linear bending compliance for any \(N> 1\) has a peak at a positive finite value of $\epsilon$, as shown in FIG.~\ref{fig:comp_n}.



Consider the large force regime, where \(f \gg k_BT/2b\) and \(f \gg \epsilon/2b\). Then, the microstate with the longest rod, {\it i.e.} \(Z(2)(0,1)\), dominates, and the others become negligible. This observation leads to the approximation of the mean deflection
\begin{eqnarray}
    \frac{\langle z \rangle_2}{2b} \approx 1-\frac{k_BT}{2fb}
    \label{eq:comp_rjfc_two_large}.
\end{eqnarray}
We can "guess" this response by considering only one dominant microstate. In this situation, it can be viewed as the extension of length \(2b\) rod, where the force-extension relation is given as the Langevin function \(\mathcal{L}(2fb/k_BT) \approx 1-k_BT/2fb\). In contrast, the uFJC follows
\begin{eqnarray}
    \frac{\langle z \rangle_{uFJC}}{Nb} = {\cal L}(fb/k_BT) \approx 1-\frac{k_BT}{fb}
    \label{eq:comp_ujfc_two_large}.
\end{eqnarray}
However, Eq.~(\ref{eq:comp_rjfc_two_large}) is precisely the large force limit of the second Eq.~(\ref{eq:mean_two}), which is exact.

\section{\label{sec:Recu_Rel}Recursion Relation}

It is not possible to obtain a partition function for bending in a closed form. However, it is still possible to get a useful recursion relation that allows exact numerical results for a finite number of hinges \(N\).

Consider a chain with $N-1$ reversible hinges, and we now add an extra chain segment at the free end of the chain. Then, the newly added \(N\) th hinge can be (1) opened or (2) closed. When (1), the contribution of the microstates to the partition function \(Z(N)\) is
\begin{eqnarray}
    Z(N-1)\times\frac{4\pi \sinh(f)}{f}
    \label{eq:recur_1}.
\end{eqnarray}
If (2), we have to consider more possible cases. Again, the \(N-1\) th hinge can be (3) opened or (4) closed. When (3), the contribution is
\begin{eqnarray}
    Z(N-2)\times e^\epsilon\frac{4\pi \sinh(2f)}{2f}
    \label{eq:recur_3}.
\end{eqnarray}
Repeating this process, we obtain
\begin{eqnarray}
    Z(N)&=&Z(N-1)\frac{4\pi \sinh(f)}{f}+Z(N-2)e^\epsilon\frac{4\pi \sinh(2f)}{2f}+ \dots \nonumber \\ 
    &+&Z(0)e^{(N-1)\epsilon}\frac{4\pi \sinh(Nf)}{Nf}+e^{N\epsilon}
    \label{eq:recur_list}.
\end{eqnarray}
Notice that the first segment is grafted. Thus, there is no contribution of the force when all hinges are closed. Therefore, the last term only contains the contribution of \(\epsilon\). Also, if \(N = 0\), there is only one grafted segment, so we have the initial condition as \(Z(0) = 1\). We could now rewrite Eq.~(\ref{eq:recur_list}) as a sum and we finally obtain
\begin{eqnarray}
    Z(N)=\sum_{j=1}^{N} e^{(j-1)\epsilon}\frac{4\pi \sinh(jf)}{jf} Z(N-j) + e^{N\epsilon}
    \label{eq:recur}.
\end{eqnarray}
For the case of \(N=2\), since we have \(Z(1)=e^{\epsilon}+4\pi \sinh(f)/f\) and \(Z(0)=1\), it can be confirmed that we recover Eq.~(\ref{eq:leverarm}).

Using the recursion relation, we could obtain an analytic expression of  \(\langle n \rangle\) and \(\langle z \rangle\) at two limits. Noting that the leading order of \(f\) in the partition function is
\begin{eqnarray}
    Z(N) = \mathcal{O}\Biggr(e^{(N-1)\epsilon}\frac{4\pi\sinh(Nf)}{Nf}\Biggr)
    \label{eq:highest_order_partit},
\end{eqnarray}
then it leads to
\begin{eqnarray}
    \lim_{f\to\infty} \langle n \rangle_N &=& \frac{N-1}{N}, \nonumber \\
    \lim_{f\to\infty} \langle z \rangle_N &=& Nb
    \label{eq:meanocc_meanext_limit}.
\end{eqnarray}
Also, considering only the leading order of the partition function, when \(f \gg k_BT/(Nb)\) and \(f \gg \epsilon/(Nb)\), it leads to the mean deflection
\begin{eqnarray}
    \frac{\langle z \rangle_N}{Nb} \approx {\cal L}(Nfb/k_BT) \approx 1-\frac{k_BT}{Nfb}
    \label{eq:def_limit}.
\end{eqnarray}
Notice that this is valid for only finite systems, where there may be a giant single rod (cluster), and all of the \(N\) segments are linked together when the force is sufficiently large. In the thermodynamic limit, Eq.~(\ref{eq:def_limit}) may not hold since we do not have a giant rod in the rFJC, which will become more clear in Sections~\ref{sec:Continuum} and~\ref{sec:Generating_Function}.

We could also obtain a closed-form expression of the partition function if the force is not applied as
\begin{eqnarray}
    Z(N) = (e^\epsilon+4\pi)^N
    \label{eq:partit_no_force}.
\end{eqnarray}
The result above is what we expected from the statistical independence of the $N$ reversible hinges, and it immediately gives
\begin{eqnarray}
    \langle n \rangle_{N,\:f=0} &=& \frac{e^\epsilon}{e^\epsilon+4\pi}, \nonumber \\
    \langle z \rangle_{N,\:f=0} &=& 0
    \label{eq:meanocc_meanext_0}.
\end{eqnarray}



\begin{figure}
    \includegraphics[width=0.48\textwidth]{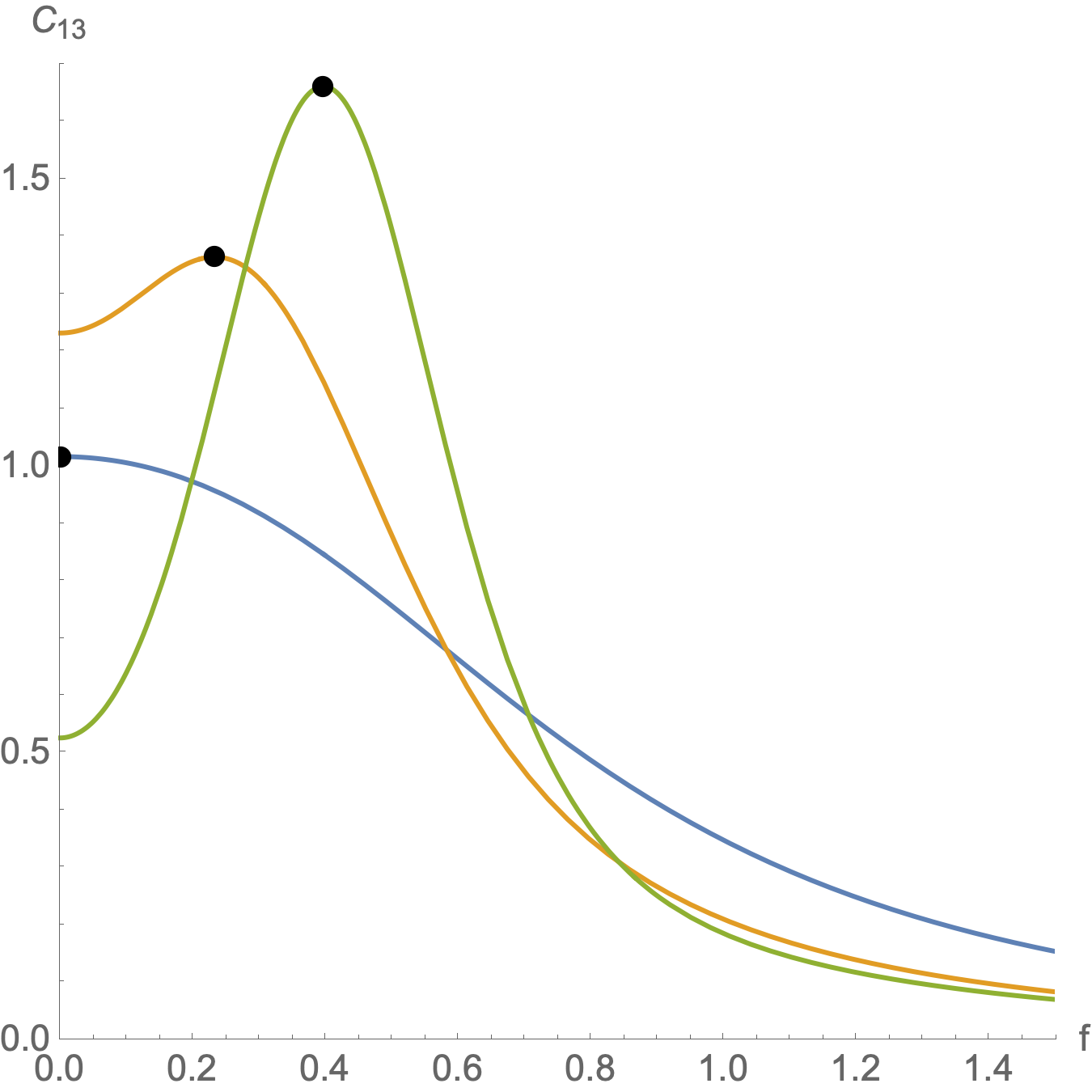}
    \caption{\label{fig:comp13}
    The bending compliance as a function of a force. The blue, yellow, and green curve refers to \(\epsilon = 3\), \(\epsilon = 4.5\), \(\epsilon = 6\), and the black dots indicate the maximum in each curve. For \(N = 13\), when \(\epsilon \gtrapprox 4\), a maximum in the compliance exists. \(N = 13\), \(k_BT = 1\), \(b = 1\). }
\end{figure}

\begin{figure}
    \includegraphics[width=0.48\textwidth]{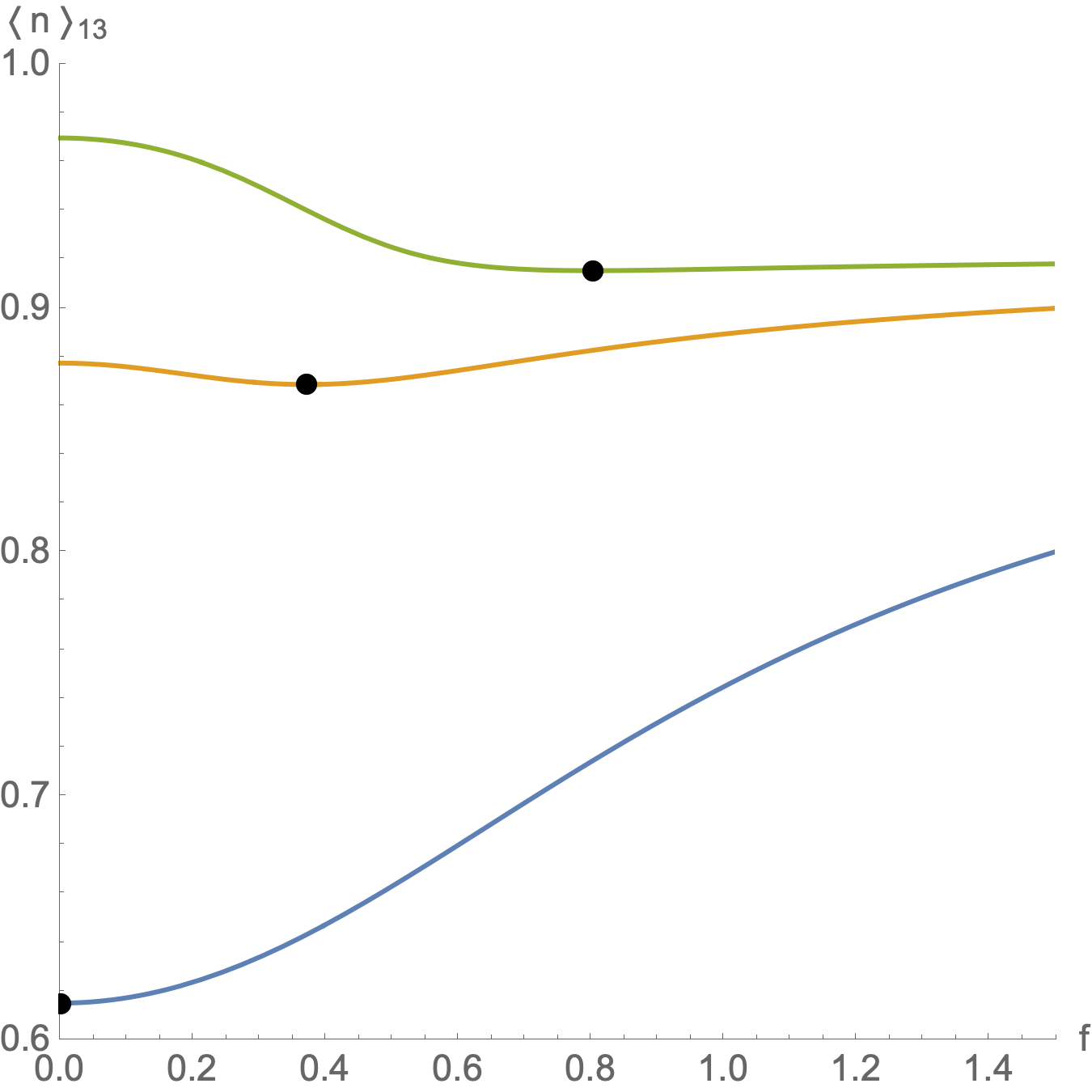}
    \caption{\label{fig:occ13}
    The mean occupation number as a function of a force. The blue, yellow, and green curve indicates \(\epsilon = 3\), \(\epsilon = 4.5\), \(\epsilon = 6\), and the black dots indicate the minimum in each curve. For \(N = 13\), when \(\epsilon \gtrapprox 4\), a maximum in the compliance and a minimum in the mean occupation number exists. \(N = 13\), \(k_BT = 1\), \(b = 1\). }
\end{figure}

As described in the head of this Section, the recursion relation allows exact numerical results for some variables, such as \(\langle n \rangle = \partial_\epsilon \ln(Z(N))/N\) and \(\langle z \rangle = \partial_f \ln(Z(N))\). Moreover, when we plot them as a function of a force, some variables can be non-monotonic. Those are the differential bending compliance \(C\) and the mean occupation number \(\langle n \rangle\). For the bending compliance, we have a maximum after some value of \(\epsilon_0\), and we have a minimum for the mean occupation number. Remarkably, those two variables start to exhibit non-monotonic behavior simultaneously. This is shown in FIG.~\ref{fig:comp13} and FIG.~\ref{fig:occ13}. Notice that the condition \(\epsilon > \epsilon_0 \approx 4\) in FIG.~\ref{fig:comp13} and FIG.~\ref{fig:occ13} is not special. When \(N\) increases, \(\epsilon_0\) also increases.

It is also interesting to point out that the global minimum in the mean occupation number \(\langle n \rangle\) shifts toward a larger value of \(f\) when the activation energy \(\epsilon\) is increased. We cannot calculate numerically if the global minimum disappears at a larger value of \(\epsilon\) or converges to a specific value of \(f\) due to the computational precision limit. However, there is a lower bound of \(\epsilon\) to observe the global minimum. A similar trend is also observed for the maximum in the bending compliance.

\section{\label{sec:Mean_Field}Mean Field Approximation}
The recursion relation in Eq.~(\ref{eq:recur}) enables us to obtain the exact partition function for a finite number of hinges. But still, it cannot be written in closed form. In this section, we make an approximation in a mean-field fashion to analyze the chain's behavior for \(N \gg 1\).

\begin{figure}
    \includegraphics[width=0.48\textwidth]{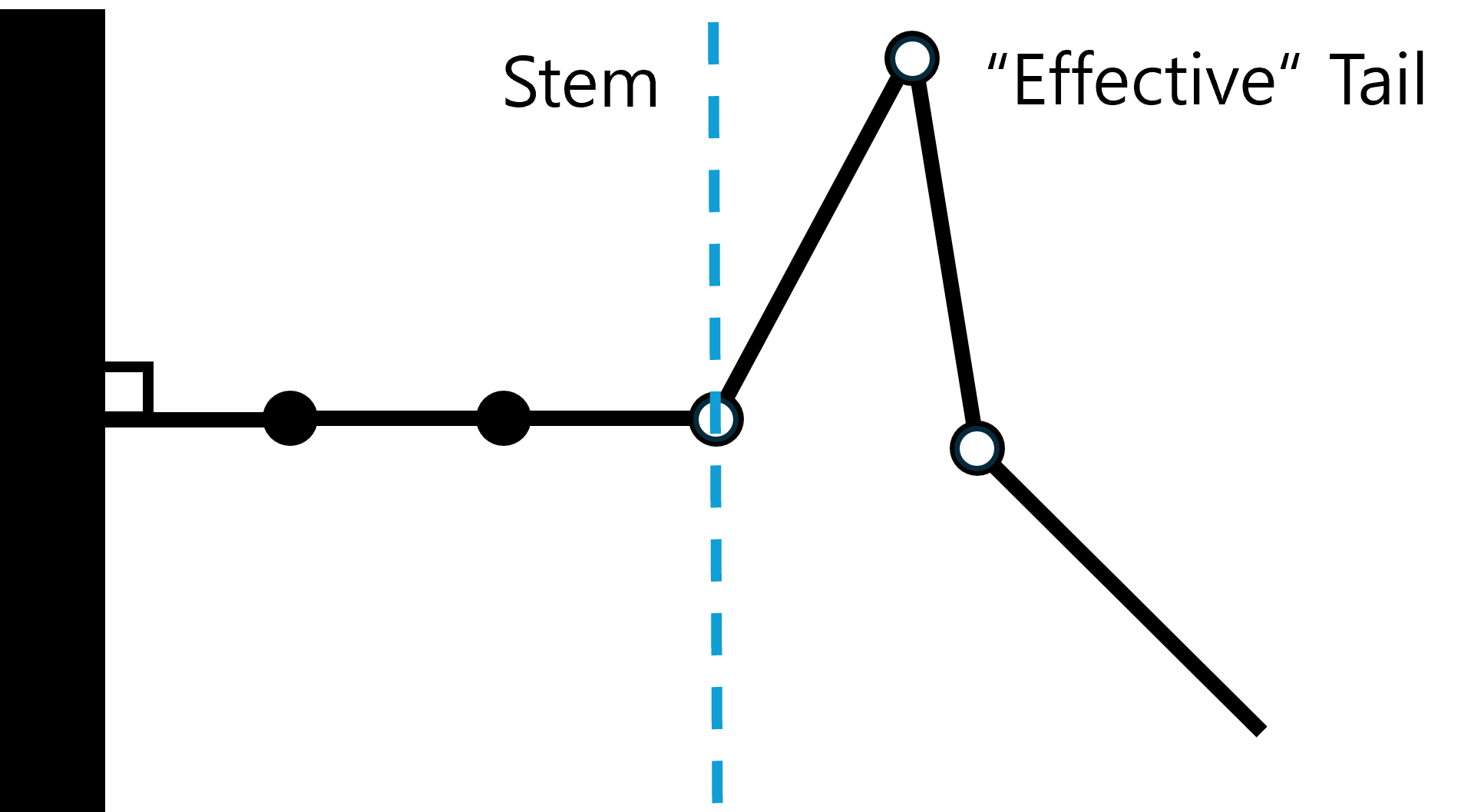}
    \caption{\label{fig:stem_tail}
    A brief illustration of how the mean field approximation is done for bending. The blue dashed line separates the stem and the (effective) tail. The tail consists of 6 segments with 2 hinges opened, and this is approximated as an effective tail with effective monomer length \(\Bar{b}=6/(2+1)b\).
    }
\end{figure}

We separate the system into two parts: the stem and the tail. We define the stem as a rod connected without being broken (opened) from the first grafted segment and the tail as the rest. For example, in FIG.~\ref{fig:stem_tail}, we assume that the chain consists of 8 hinges, and the first and second hinges are closed. Then, the stem part contains the grafted segment \(i = 1\) to \(i = 3\), and the tail contains \(i = 4\) to \(i = 9\).

In this way, the partition function can be written as
\begin{eqnarray}
    Z(N) &=& \sum_{N_S=0}^{N} \Biggr ( Z_S(N_S)\times Z_T(N-N_S-1) \Biggr ) \nonumber \\
    &=& \sum_{N_S=0}^{N} \Biggr (e^{N_S \epsilon}\times Z_T(N-N_S-1) \Biggr )
    \label{eq:MF_ZSZT},
\end{eqnarray}
where \(Z_S\) is the partition function of the stem, \(N_S\) is the number of (closed) hinges in the stem, and \(Z_T\) is the partition function of the tail.

We now implement the mean field approximation. Noticing that the tail part is just stretching an rFJC with \(N - N_S - 1\) hinges as in Ref.~\onlinecite{StretchingrFJC_gibbs}, we can use the expression of the mean-field partition function used in stretching for the Gibbs ensemble (fixed force ensemble). This way, we ignore the fluctuations of the length of the connected segments, and we view the tail as an effective uFJC. Then, the approximated partition function for bending reads
\begin{eqnarray}
    Z(N) \approx \sum_{N_S=0}^{N} e^{N_S\epsilon}
    \sum_{N_T=0}^{N-N_S-1} \Biggr[ \binom{N-N_S-1}{N_T} \times \nonumber \\ 
    e^{N_T\epsilon}\Biggr [ \frac{4\pi \sinh(\overline{b} f)}{\overline{b} f} \Biggr ]^{N-N_S-N_T} \Biggr]
    \label{eq:MF_bending},
\end{eqnarray}
where \(N_T\) is the number of closed hinges in the tail, and \(\overline{b}:= (N-N_S)b/(N-N_S-N_T)\) is the effective rod length in the tail.


\begin{figure}
    \includegraphics[width=0.48\textwidth]{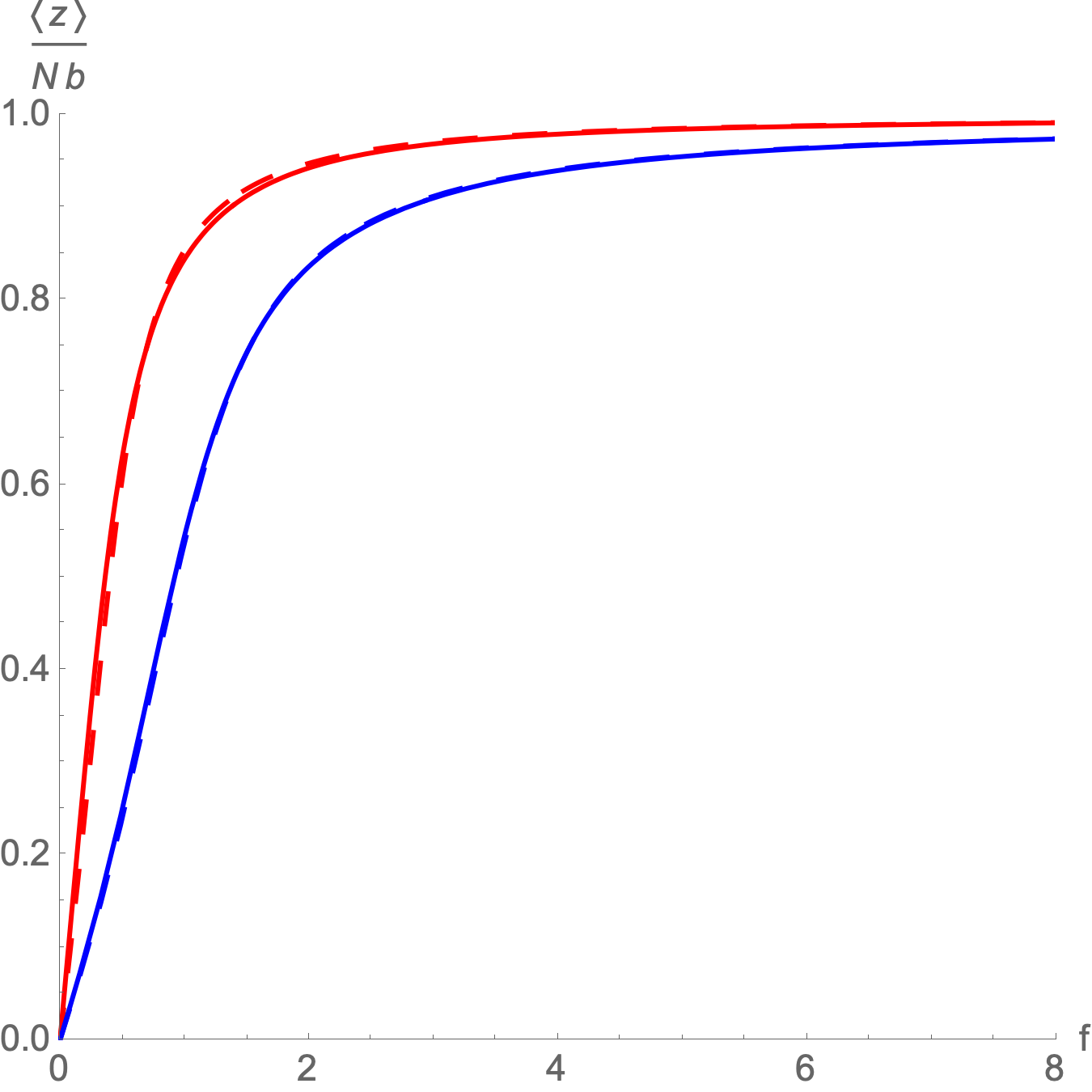}
    \caption{\label{fig:MF_ext}
    The red curve indicates the \(N = 15\) exact value of mean deflection divided by the contour length, and the red dashed curve indicates the \(N = 15\) approximate value. The blue and blue dashed curves indicate the \(N = 5\) cases. \(\epsilon = 4\), \(k_BT = 1\), \(b = 1\).
    }
\end{figure}

\begin{figure}
    \includegraphics[width=0.48\textwidth]{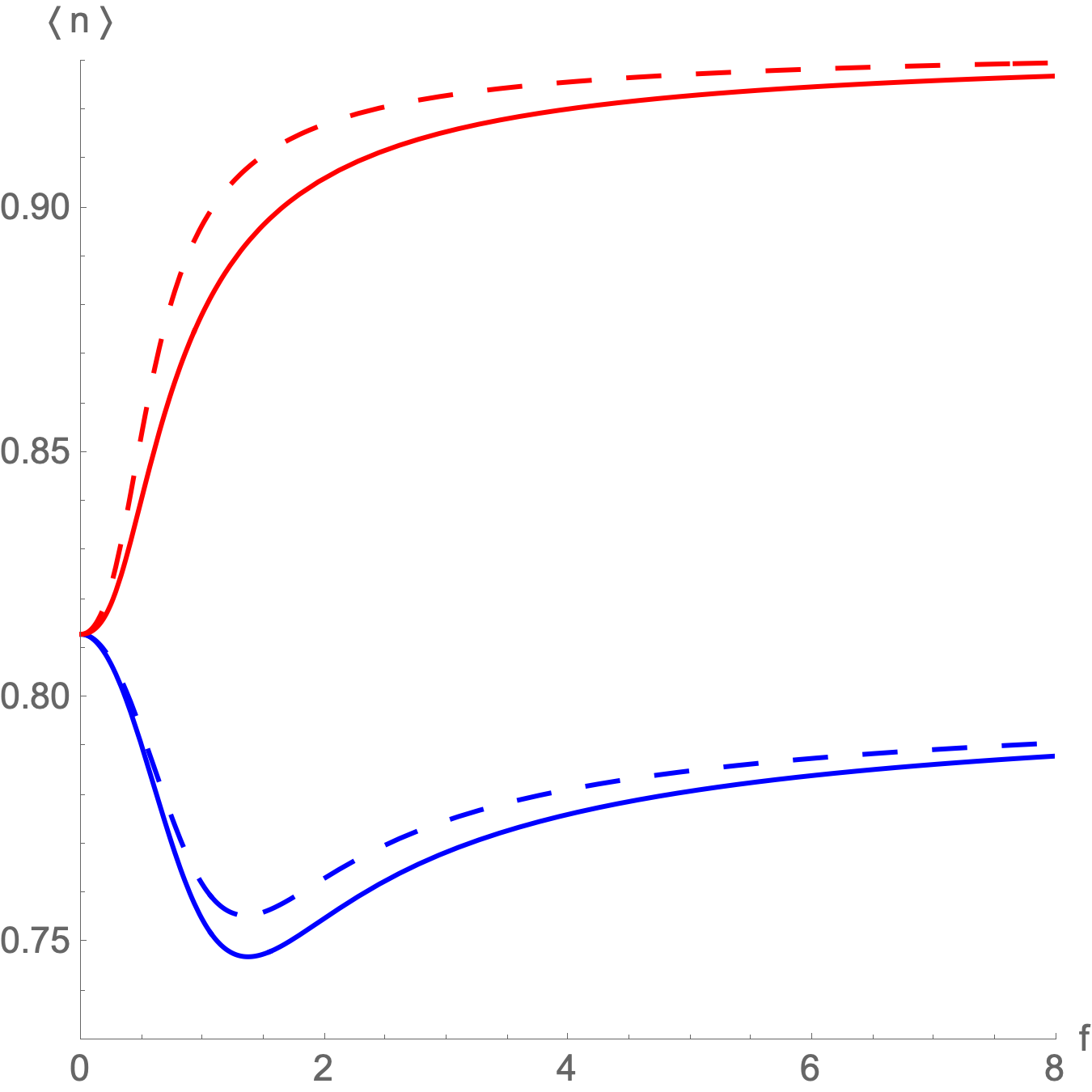}
    \caption{\label{fig:MF_occ}
    The red curve indicates the \(N = 15\) exact value of the mean occupation number, and the red dashed curve indicates the \(N = 15\) approximate value. The blue curve and the blue dashed curve are for \(N = 5\). \(\epsilon = 4\), \(k_BT = 1\), \(b = 1\).
    }
\end{figure}

Nevertheless, Eq.~(\ref{eq:MF_bending}) is still not a closed-form expression. However, it allows us to calculate the force-deflection and the mean occupation number relation more efficiently. It reduces the number of terms compared to Eq.~(\ref{eq:recur}), decreased from \(\mathcal{O}(2^N)\) to \(\mathcal{O}(N^2)\).

It also reproduces the exact value well. The comparison of the exact value and the approximated value of \(\langle z \rangle\) and \(\langle n \rangle\) is shown in FIG.~\ref{fig:MF_ext} and FIG.~\ref{fig:MF_occ}, respectively. The mean field approximation reproduces \(\langle z \rangle\) very accurately but always gives a larger value than the exact value for \(\langle n \rangle\).

\begin{figure}
    \includegraphics[width=0.48\textwidth]{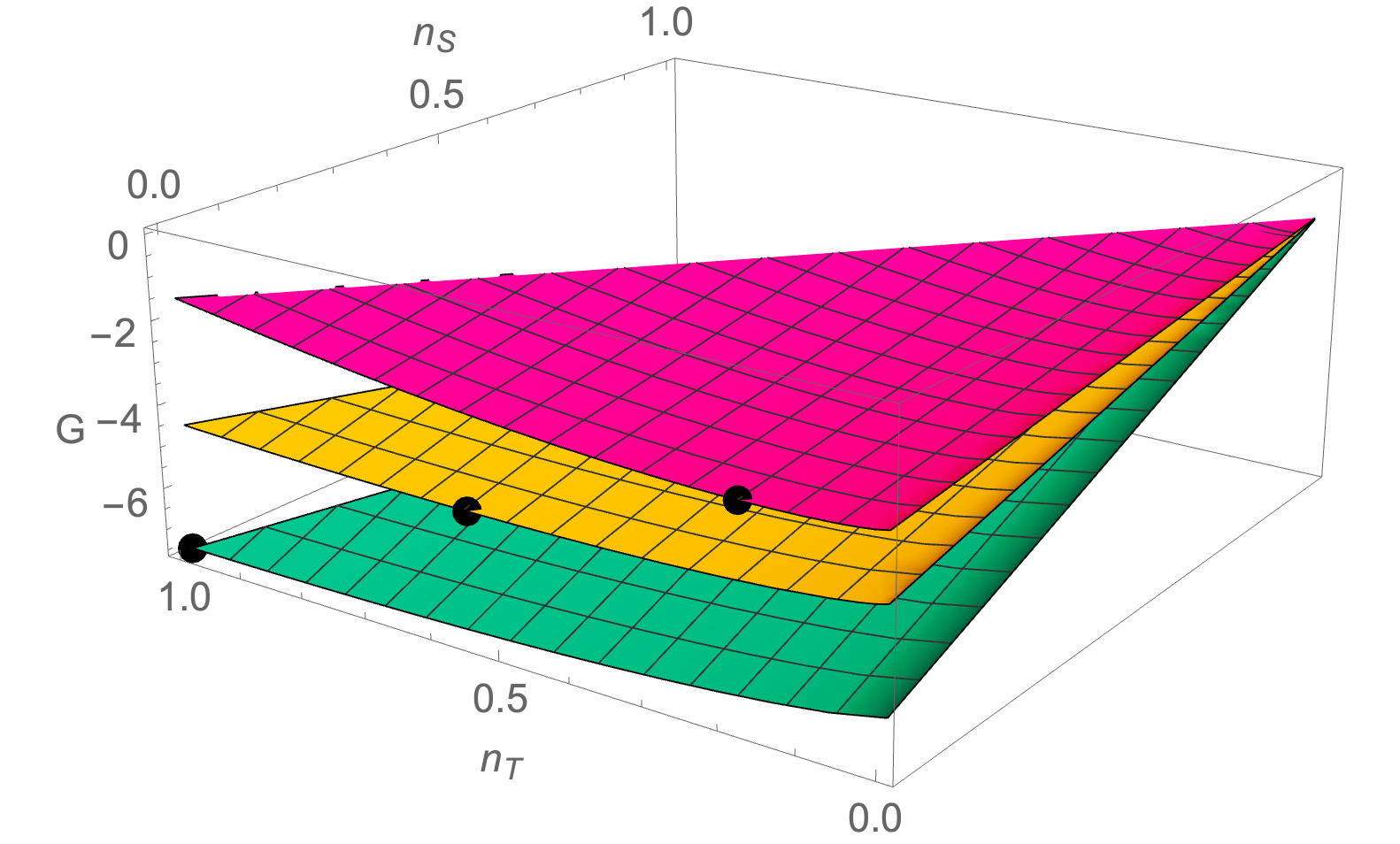}
    \caption{\label{fig:free_energy}
    The Landau free energy density is a function of two variables \(n_S\) and \(n_T\). Each curved surface represents \(\ f = 0.5\), \(3.5\), \(6.5\) (Magenta, Yellow, Cyan). The black dot on each surface indicates the global minimum. \(\epsilon = 1\), \(k_BT = 1\), \(b = 1\).
    }
\end{figure}

The approximate partition function of Eq.~(\ref{eq:MF_bending}) allows us to construct a Landau free energy density for the bending rFJC. The free energy density is a function of two variables, \(n_S=N_S/N\) and \(n_T=N_T/N\), contributing to the mean occupation number of the stem and the tail, respectively. Using the Stirling approximation, the Gibbs free energy \(G\) per segment reads
\begin{eqnarray}
    \frac{G}{N} &\approx& -(n_S+n_T)\epsilon + [(1-n_S-n_T)\ln(1-n_S-n_T) \nonumber \\
    &+& n_T\ln n_T - (1-n_S)\ln(1-n_S)] \nonumber \\
    &-&(1-n_S-n_T) \nonumber \\
    &\times&\ln \biggr[ \frac{4\pi(1-n_S-n_T)}{fb(1-n_S)}\sinh\biggr(\frac{fb(1-n_S)}{1-n_S-n_T}\biggr) \biggr]
    \label{eq:free_energy}.
\end{eqnarray}
The first term is the contribution from the activation energy of closed hinges, the second term is the contribution from the entropy of mixing, and the last term is the contribution of entropic stretching of the tail. FIG.~\ref{fig:free_energy} is the plot of the Gibbs free energy density.

Notice that the free energy density is always minimized when \(n_S = 0\) for any value of \(n_T\). This means that the stem part is negligible in the thermodynamic limit. We can say that the chain forgets its boundary condition, and it becomes the same as a chain under stretching in the thermodynamic limit. In addition, note that substituting \(n_S=0\) into Eq.~(\ref{eq:free_energy}) yields
\begin{eqnarray}
    \frac{G}{N} &\approx& -n_T\epsilon+[(1-n_T)\ln(1-n_T)+n_T\ln(n_T)] \nonumber \\
    &-& (1-n_T)\ln \biggr[ \frac{4\pi(1-n_T)}{fb}\sinh\biggr(\frac{fb}{1-n_T}\biggr) \biggr]
    \label{eq:free_energy_stretching},
\end{eqnarray}
which is the free energy density for stretching rFJC. Also, it is already known that there is no phase transition for stretching rFJC.\cite{StretchingrFJC_gibbs} Thus, we also do not expect a phase transition for bending either.

\section{\label{sec:Continuum}Continuum limit}
The continuum limit is achieved by setting \(N \rightarrow \infty\), \(b \rightarrow 0\) but keeping both the total contour length \(L = Nb\) and the persistence length \(L_p\) finite. For example, the Wormlike chain (WLC) is defined as a continuum limit of a KP chain or a Freely Rotating chain.\cite{KP_original,Wall_FRC,Rubinstein_polymer} We expect  the continuum limit to be possible for the case of rFJC as well. In this section, we obtain the continuum limit of the rFJC and analyze it.

\subsection{\label{sec:Persis_Len}Discrete Model}
The persistence length quantifies the bending stiffness of the chain. First, we calculate the persistence length for the discrete case and then we take the continuum limit, as done in the WLC.\cite{Yamakawa_helical}

First, consider the discrete scenario. If the bending force is not applied, the rFJC can be mapped to a one-dimensional percolation problem. We can think of it as a percolation problem with a uniform occupation probability \(p\) of the hinge, which is given by \(p = e^\epsilon/(e^\epsilon+4\pi)\). Then, the tangent vector auto-correlation function of the discrete rFJC with arbitrary \(N\) reads
\begin{eqnarray}
    \langle \vec{u}(i) \cdot \vec{u}(j) \rangle= \langle \cos\gamma \rangle = e^{-|i-j|b/S_p}= p^{-|i-j|}
    \label{eq:tancorr_func},
\end{eqnarray}
where \(\gamma\) is the angle made between \(\vec{u}(i)\) and \(\vec{u}(j)\).

Eq.~(\ref{eq:tancorr_func}), using a simple algebra, gives the mean squared end-to-end vector \(\langle R^2 \rangle\) and the persistence length \(S_p\) as
\begin{eqnarray}
    \langle R^2 \rangle &=& Nb^2\frac{1+p}{1-p}-2b^2p\frac{1-p^N}{(1-p)^2}
    \label{eq:endtoend_dis}, \\
    S_p &=& - b \frac{1}{\ln p}
    \label{eq:persis_length}.
\end{eqnarray}
Notice that if we replace \(p\) by \(\cos\theta_{B}\), where \(\theta_{B}\) is the bond angle in the Freely Rotating Chain model, all the expressions in Eq.~(\ref{eq:tancorr_func}) to Eq.~(\ref{eq:R_squared}) are the same as those for the Freely Rotating chain, which is known as one of the discrete versions of the Wormlike chain.\cite{KP_original,Rubinstein_polymer,Yamakawa_helical}


\subsection{\label{sec:Equations}Basic equations and Persistence length}
The Brownian motion of a particle can be modeled as a Markov process in 6-dimensional phase space.\cite{Chandrasekhar_Stochastic, Livi_Noneq} This is also true for our model (in the absence of force) since we are assuming local short-range correlations, but now the polymers diffuse in length.

Let \(\vec{R}\) and \(\vec{u}\) be the end-to-end vector and the unit tangent vector of the last segment of the chain, respectively. Then, the Chapman-Kolmogorov equation is written as
\begin{eqnarray}
    W(\vec{R},\vec{u}|\vec{u_0};L+b)=\iint W(\vec{R}-\Delta\vec{R},\vec{u}-\Delta\vec{u}|\vec{u_0};L)\times \nonumber \\
    \Psi(\Delta\vec{R},\Delta\vec{u}|\vec{R}-\Delta\vec{R},\vec{u}-\Delta\vec{u}) \,d(\Delta\vec{R}) \,d(\Delta\vec{u})
    \label{eq:standard_meq},
\end{eqnarray}
where \(W(\vec{R},\vec{u}|\vec{u_0};L)\) is the joint (positional-orientational) probability distribution of the last segment when the initial tangent vector is given by \(\vec{u_0}\) with the constraint of contour length \(L\), and \(\Psi(\Delta\vec{R},\Delta\vec{u} |\vec{R}-\Delta\vec{R},\vec{u}-\Delta\vec{u})\) is the propagator (transition probability).

To obtain the corresponding Fokker-Planck equation, a standard procedure requires expanding Eq.~(\ref{eq:standard_meq}) to the first order of \(b\). However, this can not be done for the rFJC, since the leading order of \( \langle \Delta\vec{u}^2 \rangle \) is the second order of b. To overcome this problem, we change the variable as \(\Delta\vec{u}=\vec{u}-\vec{u}'\). Then, Eq.~(\ref{eq:standard_meq}) can be written as
\begin{eqnarray}
    W(\vec{R},\vec{u}|\vec{u_0};L+b)=\iint W(\vec{R}-\Delta\vec{R},\vec{u}'|\vec{u_0};L)\times \nonumber \\
    \Psi(\Delta\vec{R},\vec{u}|\vec{R}-\Delta\vec{R},\vec{u}') \,d(\Delta\vec{R}) \,d\vec{u}',
    \label{eq:change_var_meq}
\end{eqnarray}
where
\begin{eqnarray}
    \Psi(\Delta\vec{R},\vec{u}|\vec{R}-\Delta\vec{R},\vec{u}') = \delta(\Delta\vec{R}-b\vec{u})\psi(\vec{u}|\vec{u}')
    \label{eq:inext_cond}.
\end{eqnarray}
Then, Eq.~(\ref{eq:change_var_meq}) reduces to
\begin{eqnarray}
    W(\vec{R},\vec{u}|\vec{u_0};L+b)=\int W(\vec{R}-b\vec{u},\vec{u}'|\vec{u_0};L) \psi(\vec{u}|\vec{u}')\,d\vec{u}',
    \label{eq:rFJC_meq}
\end{eqnarray}
with
\begin{eqnarray}
    \psi(\vec{u}|\vec{u}') = p\delta(\vec{u}'-\vec{u})+(1-p)\frac{1}{4\pi}
    \label{eq:propagator}.
\end{eqnarray}
Notice that \(p\) is the probability of the hinge to be closed.

Finally, we obtain the basic differential equation for rFJC by expanding Eq.~(\ref{eq:rFJC_meq}) to the first order of \(b\) as
\begin{eqnarray}
    \Biggr(\frac{\partial}{\partial L}+\vec{u}\cdot\grad_R \Biggr) W(\vec{R},\vec{u}|\vec{u_0};L)= & \nonumber \\
    -\frac{1}{L_p}W(\vec{R},\vec{u}|\vec{u_0};L)+\frac{1}{L_p}\frac{1}{4\pi} \int W(\vec{R},\vec{u}'|\vec{u_0};L) \,d\vec{u}'
    \label{eq:rFJC_DE},
\end{eqnarray}
Where we have set \(L_p=b/(1-p)\) and the local inextensibility constraint \(|\vec{u}|=1\). Notice that the parentheses in the LHS of the equation above are the convective derivative along the contour (path of the particle) \(\vec{R}\) with the tangent vector (velocity) \(\vec{u}\). The equation above leads to the differential equation for \(W(\vec{u}|\vec{u_0};L)\) as
\begin{eqnarray}
    \frac{\partial W(\vec{u}|\vec{u_0};L)}{\partial L}=-\frac{1}{L_p}W(\vec{u}|\vec{u_0};L)+\frac{1}{L_p}\frac{1}{4\pi},
    \label{eq:rFJC_tangent_DE}
\end{eqnarray}
by taking its Fourier transform  with respect to \(\vec{R}\).

The solution of Eq.~(\ref{eq:rFJC_tangent_DE}) is easily obtained. Noting that the initial condition is given as \(W(\vec{u}|\vec{u_0};0)=\delta(\vec{u}-\vec{u_0})\), the exact solution is
\begin{eqnarray}
    W(\vec{u}|\vec{u_0};L)=\frac{1}{4\pi}(1-e^{-L/L_p}) + \delta(\vec{u}-\vec{u_0})e^{-L/L_p}
    \label{eq:PDF_tangent}.
\end{eqnarray}
The solution can be intuitively understood. If the hinges are all closed, the chain should be oriented in the same direction as the initial tangent vector. This refers to the second term in the RHS. However, the last segment can be oriented freely if at least one hinge is opened. Thus, for the latter, the tangent vector distribution is uniform along the unit sphere, which refers to the first term.

Now, let us consider the persistence length \(P\) at the continuum limit, which is defined as
\begin{eqnarray}
    \langle \vec{u}_0 \cdot \vec{u} \rangle= \langle \cos \gamma \rangle = e^{-L/P}
    \label{eq:cont_persis_def}.
\end{eqnarray}
Note that Eq.~(\ref{eq:PDF_tangent}) leads to the persistence length
\begin{eqnarray}
    \int (\vec{u_0}\cdot\vec{u}) W(\vec{u}|\vec{u_0};L) \,d\vec{u} = e^{-L/L_p}
    \label{eq:cont_persis},
\end{eqnarray}
implying that the persistence length is \(P=L_p=b/(1-p)\) in the continuum limit. Then, it follows that
\begin{eqnarray}
    \langle R^2 \rangle = 2L_pL - 2L_p^2(1-e^{-L/L_p})
    \label{eq:R_squared},
\end{eqnarray}
which is the same expression as in the WLC, since in both models we have an exponential decay of the orientational correlation and the local inextensibility constraint.

\subsection{\label{sec:Moments}Moments}
According to linear response theory, quadratic moments of the undisturbed system determine the initial linear response. In this Section, we obtain the moments \(\langle \theta^m \rangle\) and \(\langle z^2 \rangle\), in the absence of an applied force.

Only in this Section, we define \(\theta\) as an angle between the \(\hat{x}\) axis (the tangent vector of the initial grafted segment) and the last segment, which is the bending angle. To obtain \(\langle \theta^m \rangle\), we need the PDF of the orientation, and this is done in Eq.~(\ref{eq:PDF_tangent}). By using proper notations under the constraint \(|\vec{u}|=1\) in spherical coordinates, the PDF of the orientation reads
\begin{eqnarray}
    W(\vec{u}|\vec{u_0};L)=&=&\frac{1}{4\pi}(1-e^{-L/L_p}) \nonumber \\ 
    &+& \frac{\delta(\theta-\theta_0)\delta(\phi-\phi_0)}{\sin\theta}e^{-L/L_p}. \quad
    \label{eq:PDF_omega}
\end{eqnarray}
Noting that \(\theta_0=0\) when \(\vec{u_0}=\hat{x}\), the moment \(\langle \theta^m \rangle\) is
\begin{eqnarray}
    \langle \theta^m \rangle = \frac{1}{2}(1-e^{-L/L_p})
    \int_{0}^{\pi} \theta^m \sin\theta\, d\theta
    \label{eq:moment_theta}.
\end{eqnarray}
For example, for \(m=2\), \(\langle \theta^2 \rangle = (1-e^{-L/L_p})(\pi^2-4)/2\).

Let us now focus on \(\langle z^2 \rangle\). For a flexible chain, \(\langle z^2 \rangle = \langle R^2 \rangle/3 \) due to isotropy. However, if \(L \lessapprox L_p\), the grafted boundary condition affects the last segment, and it follows that  \(\langle z^2 \rangle \neq \langle R^2 \rangle/3\). Nonetheless, finding an approximate expression in the rod-like limit \(L \ll L_p\) is still possible.

We aim to obtain an expression to the first order in \(L/L_p\). We proceed by considering all possible cases that give a significant contribution to \(\langle z^2 \rangle\). Imagine a chain with \(N\) hinges. At least one hinge should be opened to contribute to \(\langle z^2 \rangle\). First, think about the case when only one hinge is opened. Then the contribution to \(\langle z^2 \rangle\) for this case is
\begin{eqnarray}
    (1-p)p^{N-1}\Biggr[\frac{N^2 b^2}{3} + \frac{(N-1)^2 b^2}{3} + \dots +  \frac{1^2 b^2}{3} \Biggr]
    \label{eq:z_squared_first_sum}.
\end{eqnarray}
The factor \((1-p)p^{N-1}\) refers to the probability that only one hinge is opened, and the terms in parentheses refer to the value of \(z^2\), when the first, second, and the \(N\)th hinge is opened. Noting that \(1-p = b/L_p\), and taking the continuum limit, it turns out that the leading order of the expression in Eq.~(\ref{eq:z_squared_first_sum}) is
\begin{eqnarray}
    \frac{L^3}{9L_p}
    \label{eq:z_squared_first}.
\end{eqnarray}
Now, consider the case when two hinges are opened. Then the contribution to \(\langle z^2 \rangle\) is
\begin{eqnarray}
    (1-p)^2p^{N-2}\mathcal{O}(N^4 b^2) = \mathcal{O}\Biggl(\frac{L^4}{L_p^2}\Biggr)
    \label{eq:z_squared_second_sum}.
\end{eqnarray}
We see that this is negligible, in this limit. Then, it follows that the case of more opened hinges is also negligible in the rod-like limit. Thus, we conclude that 
\begin{eqnarray}
\langle z^2 \rangle \approx \frac{L^3}{9L_p}\;,
\end{eqnarray}
when \(L \ll L_p\).

\subsection{\label{sec:Rod_Lim}Rod-like Limit}
The rod-like limit is when \(L \ll L_p\) in the continuum limit, keeping \(L_p/L = \alpha\) as a constant where \(\alpha \gg 1\). This means that the orientational correlation has not vanished at the end of the chain. In this Section, we aim to calculate the bending compliance and the mean occupation number as a function of a bending force in the small force regime \(f \ll k_BT/(Nb)\) by expanding the exact partition function Eq.~(\ref{eq:recur}) in terms of \(f\) in the limit of \(N \rightarrow \infty\), keeping $L=Nb$ fixed.

The calculation of the force-deflection relation is done in Appendix.~\ref{sec:ty_appendix}. The resulting force-deflection relation, Eq.~(\ref{eq:meanext_3}), leads to the bending compliance
\begin{eqnarray}
    C=\frac{1}{L}\frac{\partial\langle z \rangle}{\partial f}&=& \frac{L}{k_B T} \Biggr[ \frac{L}{9 L_p}-\frac{L^2}{18 L_p^2}+ \mathcal{O}\Biggr(\frac{L^3}{L_p^3} \Biggr) \Biggr]+\nonumber \\
    &&\frac{f^2L^3}{k_B^3T^3}\Biggr[\frac{L}{50 L_p} + \mathcal{O}\Biggr(\frac{L^2}{L_p^2} \Biggr) \Biggr]
    \label{eq:comp_stiff}.
\end{eqnarray}
This is what we would expect from linear response theory, where the leading order of the differential compliance in the small force regime is proportional to Eq.~(\ref{eq:z_squared_first}), the \(\langle z^2 \rangle\). The mean occupation number \(\langle n \rangle\) can also be obtained following the process shown in Appendix.~\ref{sec:ty_appendix}. The mean occupation number in the rod-like and small force limit reads
\begin{eqnarray}
    \langle n \rangle \approx \frac{e^\epsilon}{e^\epsilon+4\pi}\Biggr(1-\frac{f^2}{6k_B^2T^2}\frac{L^3}{3L_p}\Biggr)
    \label{eq:meanocc_small}.
\end{eqnarray}
Noting that \(e^\epsilon/(e^\epsilon+4\pi)\) is the mean occupation number in the absence of any applied force, we see that a hinge "breaks (opens)" only when a bending force is applied in this small force regime. Eq.~(\ref{eq:meanocc_small}) also tells us that the linear response regime shrinks when \(L_p\) gets smaller because the second term in the parentheses should always be much smaller than 1.

\begin{figure}
    \begin{tikzpicture}
    \begin{axis}[
    axis lines = left,
    xlabel = \(f\),
    ylabel = {\(\langle n \rangle\)},
    xmin=0,
    ymin=0.95, ymax=1.05,
    xtick=\empty,
    ytick=\empty,
    xticklabel=\empty,
    yticklabel=\empty,
    every axis plot/.append style={thick}
]
\addplot [
    domain=0:0.5, 
    samples=500, 
    color=red,
]
{0.997 - 0.1*x^2};

\addplot [
    domain=1.5:2.5, 
    samples=1000, 
    color=blue,
]
{1 - (x^2)*exp(-x^2)/20};

\addplot[
    color=black,
    mark=\empty,
    style = dashed
    ]
    coordinates {
    (0.5,0.972)(0.6,0.967)(0.7,0.965)(0.75,0.9654)(0.8,0.966)(0.85,0.9667)(0.9,0.968)(0.95,0.9701)(1.0,0.972)(1.1,0.9765)(1.2,0.9805)(1.3,0.9835)(1.4,0.986)(1.5,0.988)
    };

\node[] at (axis cs: 2,1.005) {Asymptote \(\langle n \rangle\rightarrow 1\)};
\node[] at (axis cs: 1,1.01) {\(e^\epsilon/(4\pi+e^\epsilon)\)};
\draw[color=black, dashed] (axis cs:0,1) --  (axis cs:2.5,1);
\draw[ultra thick, color=black, dashed] (axis cs:0.5,0) -- (axis cs:0.5,1.1);
\draw[ultra thick, color=black, dashed] (axis cs:1.5,0) -- (axis cs:1.5,1.1);
\node[anchor=west] (source) at (axis cs:-0.05,0.9962){\textbullet\ };
\node (destination) at (axis cs:0.7,1.01){};
\draw[->](source)--(destination);

\end{axis}
\end{tikzpicture}
\caption{\label{fig:exist_min} An illustration about the existence of global minimum at the rod-like limit of \(\langle n \rangle\). The red curve is the plot of the small force regime (Eq.~(\ref{eq:meanocc_small})), and the blue curve is the plot of the large force regime (Eq.~(\ref{eq:thm_def_occ})). The two vertical lines indicate the intermediate region, and the dashed curve is the putative exact solution. }
\end{figure}
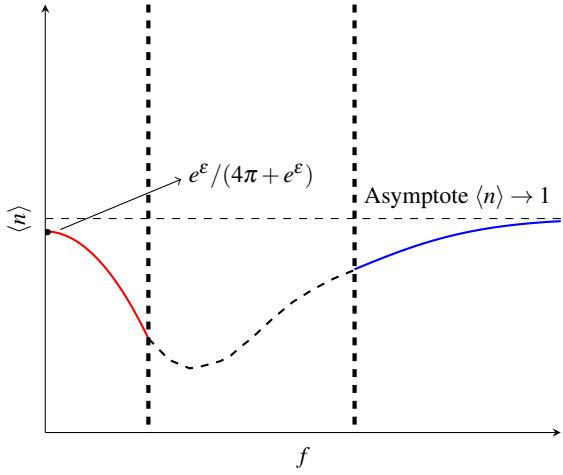

Since the mean occupation number \(\langle n \rangle\) should approach \(1\) and the bending compliance \(C\) should approach \(0\) when the bending force is increased in the large force regime (essentially becoming a stretching force), it is now clear that we have non-monotonic behavior for both quantities in the rod-like limit. Notice that, in the small force regime where Eq.~(\ref{eq:meanocc_small}) applies, the mean occupation number decreases by \(\sim f^2\). Thus, we can conclude that a global minimum in the force-mean occupation number relation exists when \(\epsilon>\epsilon_0 \approx k_BT\ln(4\pi \alpha N )\) and \(\alpha\gg1\). This is illustrated in FIG.~\ref{fig:exist_min}. In the case of the force-bending compliance relation, it increases by \(\sim f^2\) in the small-force regime, and it decreases to \(0\) in the large force regime, indicating that there exists a global maximum when \(\epsilon > \epsilon_0 \approx k_BT\ln(4\pi \alpha N)\) and \(\alpha\gg1\).

\section{\label{sec:Generating_Function}Generating Function Method \\ (Necklace Model)}

\subsection{\label{sec:General_Result}Generating Functions}

The generating function method is a mathematical technique with a broad range of applicability from graph theory and special functions in mathematics to percolation theory and field theory in physics. It is particularly useful in the statistical physics of many linear systems such as polymers. \cite{Fisher1984WalksWW,Rudnick_book,Grosberg} Here, we obtain two generating functions whose \(k\)-th coefficient is the partition function of a stretching rFJC with \(k-1\) hinges and a bending rFJC with \(k\) hinges, respectively. First, we deal with the stretching case. From Ref.~\onlinecite{StretchingrFJC_gibbs}, the partition function \(Z_S(N)\) for the stretching is given as
\begin{eqnarray}
    Z_S(\,0\,)&=&g(1), \nonumber \\
    Z_S(\,1\,)&=&g(1) Z_S(0) + g(2), \\
    Z_S(\,2\,)&=&g(1) Z_S(1) + g(2) Z_S(0) + g(3), \nonumber \\
    Z_S(N)&=&g(1) Z_S(N-1) + \dots + g(N) Z_S(0) + g(N+1) \nonumber
    \label{eq:stretch_example},
\end{eqnarray}
where \(Z_S(-1)=1\) and \(g(j) = e^{(j-1)\epsilon}4\pi \sinh(jf)/jf\). We define two generating functions. Let \(\zeta_S(x)\) and \(G(x)\)  be the generating functions whose coefficients are  \(Z_S(k-1)\) and \(g(j)\), respectively:
\begin{eqnarray}
    \zeta_S(x)&:=&\sum_{k=0}^{\infty} Z_S(k-1) x^k, \nonumber \\
    G(x)&:=&\sum_{j=1}^{\infty} g(j) x^j
    \label{eq:g_z_gf}.
\end{eqnarray}
\(\zeta_S(x)\) is the generating function of a rFJC under stretching and $G(x)$ is just an auxiliary generating function that is useful for our calculation.
Then, \(\zeta_S(x)\) can be written as
\begin{eqnarray}
    \zeta_S(x)&=&1+G(x)+G(x)^2+G(x)^3+\dots \nonumber \\
    &=&\frac{1}{1-G(x)}
    \label{eq:gf_stretching_sum}.
\end{eqnarray}
Before we proceed, we should make clear that writing the generating function as above implies the number of hinges \(N\) is strictly infinite compared to any other variables, \(\epsilon\) and \(f\), {\it i.e.} \(L \gg L_p\).\cite{brualdi2012combinatorics} (In the language of percolation theory, there is no "giant rod".) Otherwise, the geometric series must end. This point must be considered when interpreting the result.

The equation above tells us that if we know the closed-form expression of \(G(x)\), then we also know \(\zeta_S(x)\). Now, our job is to find \(G(x)\). A simple manipulation of the summation given in the second line in Eq.~(\ref{eq:g_z_gf}) yields an ODE
\begin{eqnarray}
    \frac{dG}{dx}=\frac{2\pi}{e^\epsilon f}\frac{1}{x}\Biggr( \frac{1}{1-e^{\epsilon+f}x}-\frac{1}{1-e^{\epsilon-f}x} \Biggr),
    \label{eq:g_gf_de}
\end{eqnarray}
where the initial condition is clearly \(G(0)=0\). The solution of the ODE above can be easily obtained as
\begin{eqnarray}
    G(x)=\frac{2\pi}{e^\epsilon f}\Bigl[\ln(1-e^{\epsilon-f}x)-\ln(1-e^{\epsilon+f}x)\Bigr]
    \label{eq:gf_g}.
\end{eqnarray}

A similar method can be applied to the bending case, and it yields a generating function \(\zeta(x)\) whose \(k\)-th coefficient is the partition function of a rFJC under bending with $k$ hinges, \(Z(k)\). The generating function for bending reads
\begin{eqnarray}
    \zeta(x)=\frac{1}{(1-G(x))(1-e^\epsilon x)}
    \label{eq:gf_bending}.
\end{eqnarray}
We can easily understand the form of this equation because the term $1/(1-e^\epsilon x)$ generates a geometric series that accounts for the contribution of the stem part (at the grafted end). 

\begin{figure}
\includegraphics[width=0.48\textwidth]{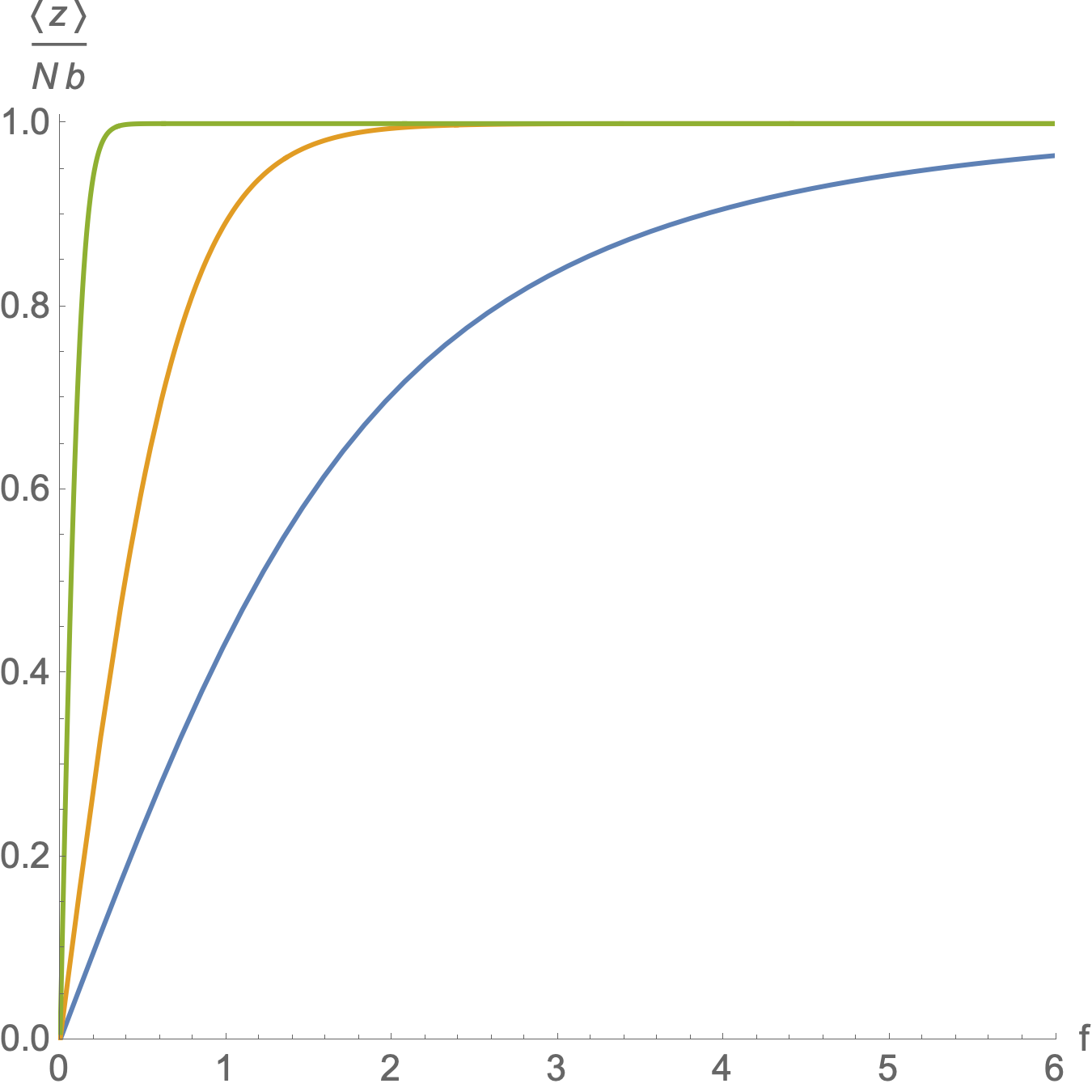}
\caption{\label{fig:gf_deflec}
The mean deflection as a function of a force in the thermodynamic limit. The blue, orange, and green curves indicate different values of \(\epsilon\), \(\epsilon=1\), \(\epsilon=3\), and \(\epsilon=5\) respectively. \(k_BT = 1\), \(b = 1\).
}
\end{figure}

\begin{figure}
\includegraphics[width=0.48\textwidth]{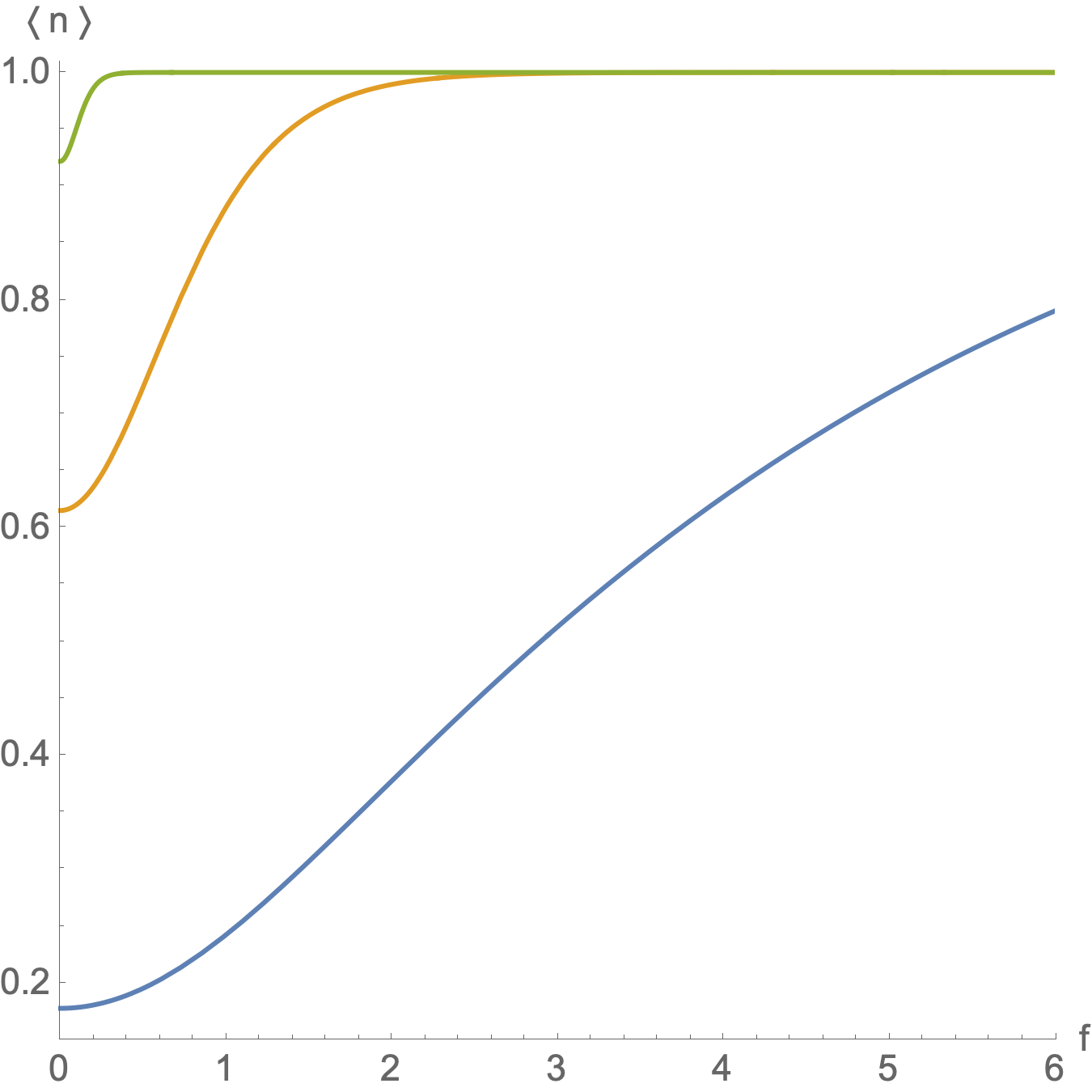}
\caption{\label{fig:gf_occ}
The mean occupation number as a function of a force in the thermodynamic limit. The blue, orange, and green curves indicate \(\epsilon=1\), \(\epsilon=3\), and \(\epsilon=5\) respectively. \(k_BT = 1\), \(b = 1\).
}
\end{figure}

The behavior of a linear system in the thermodynamic limit is dominated by the smallest real positive pole \(p^*\) of the generating function (or of the grand canonical partition function as a function of a complex fugacity). Of course, this is also true for the rFJC. The smallest real positive pole of the generating function, Eq.~(\ref{eq:gf_stretching_sum}), for stretching reads
\begin{eqnarray}
    p_1=\frac{\exp(\frac{e^\epsilon f}{2 \pi})-1}{e^{\epsilon+f}\exp(\frac{e^\epsilon f}{2 \pi})-e^{\epsilon-f}}
    \label{eq:substantial_pole}.
\end{eqnarray}
We have two poles for the bending generating function, given in Eq.~(\ref{eq:gf_bending}): \(p_1\) and an additional pole \(p_2=e^{-\epsilon}\). However, the smallest pole is still \(p^*=p_1 \leq p_2\) for all \(\epsilon\) and \(f\). Thus, we conclude that the bending is equivalent to stretching in the thermodynamic limit. Also, using the fact that \(\ln p^*\) acts as a free energy density in the thermodynamic limit, we can obtain the force-deflection and the force-mean occupation number relation. These results are shown in Appendix.~\ref{sec:gf_appendix}.

Since the smallest non-negative pole is the same in both stretching and bending, the force-extension relation and the mean occupation number are the same for both cases, in the thermodynamic limit. This can be understood by realizing that the boundary conditions are irrelevant in the thermodynamic limit, and so is the stem part. Eq.~(\ref{eq:thm_def_occ}) shows the exact stretching or the bending elasticity for arbitrary \(f\) in the thermodynamic limit. This is plotted in FIG.~\ref{fig:gf_deflec}. and ~\ref{fig:gf_occ}. Notice that we do not have non-monotonic behavior for the differential compliance or the mean occupation number.

It is interesting to examine the approximate behavior in the small force regime and the large force regime. In the small-force regime, an approximation of Eq.~(\ref{eq:thm_def_occ}) yields
\begin{eqnarray}
    \frac{\langle z \rangle}{L}&\approx& \frac{1}{3} \Biggr[\frac{1+p}{1-p}\Biggr]\frac{fb}{k_BT} \nonumber \\&-&\frac{e^{\frac{3 \epsilon }{k_B T}}+6 \pi  e^{\frac{2 \epsilon }{k_B T}}+16 \pi ^2 e^{\frac{\epsilon }{k_B T}}+16 \pi ^3}{720 \pi^3}\frac{f^3 b^3}{k_B^3 T^3},
    \label{eq:approx_def_small}
\end{eqnarray}
and
\begin{eqnarray}
    \langle n \rangle &\approx& p+\frac{e^{\frac{\epsilon}{k_BT}}}{12 \pi} \frac{f^2 b^2}{k_B^2T^2} \label{eq:approx_occ_small},
\end{eqnarray}
where \(p=e^\epsilon/(4\pi+e^\epsilon)\) is the occupation probability when \(f=0\). For the large-force regime,
\begin{eqnarray}
    \frac{\langle z \rangle}{L}&\approx& 1-\frac{e^{\frac{\epsilon}{k_BT}}}{2\pi} \exp(-\frac{f b e^{\frac{\epsilon}{k_BT}}}{2\pi k_BT}), \nonumber \\
    \langle n \rangle &\approx& 1-\left(\frac{f b e^{\frac{\epsilon}{k_BT} }}{2 \pi k_B T }+1\right) \exp \left(-\frac{f b e^{\frac{\epsilon}{k_BT}}}{\pi k_BT }\right) \nonumber \\
    &-&\frac{f b e^{\frac{\epsilon}{k_BT}}}{2 \pi k_B T }\exp \left(-\frac{f b e^{\frac{\epsilon}{k_BT}}}{2 \pi k_B T}\right)
    \label{eq:approx_large}.
\end{eqnarray}

Before we end this section, we should make a few comments on the results obtained by the generating function method. First, notice that the rFJC has qualitatively different behaviors from the uFJC in the strongly bent regime. For the mean deflection, it approaches \(1\) with an exponential decay by increasing the force. This result is very different from the corresponding result for the uFJC, which approaches \(1\) as a power law decay. Besides, the rFJC also approaches \(1\) with an exponential decay in the mean occupation number.

Let us now focus on the weakly bent regime. For the small force regime, we have written the result in terms of \(p\) to make contact with percolation theory.\cite{stauffer1992percolation} The cluster size \(S\) yielded by the 1-D percolation theory is \(S=(1+p)/(1-p)\), which we can observe in Eq.~(\ref{eq:approx_def_small}). Thus, one can easily notice that when \(N\) is strictly infinite, the rFJC can be viewed as a "random walk in diffusive regime" with a step size \(S\), which also agrees with the calculation of \(\langle R^2 \rangle\), Eq.~(\ref{eq:endtoend_dis}). Also, note that the compliance always decreases, {\it i.e.} no non-monotonic behavior appears for the compliance.

Lastly, the occupation number increases by \(f^2\). Notice that the coefficient of the higher-order term \(f^2\) is always positive. This also indicates there's no non-monotonic behavior both in the force-compliance and the force-occupation number relation, unlike the rod-like case \(L \ll L_p\).

\section{\label{sec:Conclusion}Conclusion and Discussion}

In this article, we have analyzed the behavior of the rFJC under bending in the Gibbs ensemble. At first glance, we expected that the rFJC would be less compliant than its counterpart, which is the usual freely jointed chain due to the existence of reversible hinges. However, it turns out that the lever-arm effect takes part when \(N\geq2\), and it may become more compliant than the uFJC under certain conditions.

We have derived a recursion relation, which produces the exact partition function for a finite number of hinges. This allows exact-numerical calculations for the various thermodynamic observables for a chain with a finite number of hinges. We have observed non-monotonic behavior in the force-bending compliance and the force-mean occupation number relation when \(\epsilon\) exceeds some value and the chain becomes stiffer.

Making an observation that the partition function can be broken down into two parts, the stem and the tail, and using the fact that the physical state of the tail is the same as that of the rFJC under a tensile force, the mean-field version of the partition function can be obtained. Using the mean-field version of the partition function, allowed the construction of a Landau free energy density. It turns out that the stem does not take part in the thermodynamic limit, which means it becomes negligible. Thus, bending becomes equivalent to stretching in the thermodynamic limit.

Moreover, we have considered the chain conformations. We first considered the discrete chain and calculated the tangent vector autocorrelation function of an rFJC with an arbitrary number of hinges \(N\), persistence length \(S_p\), and mean square end-to-end distance \(\langle R^2 \rangle\). Then, noting that the continuous rFJC can be viewed as a Markov process propagating in phase space along the contour length \(L\), we obtained a differential equation for the continuous rFJC and solved the PDF of the tangent vector analytically. Using this result, it turns out that the persistence length is \(L_p=b/(1-p)\) in the continuum limit. We also obtained the moments \(\langle R^2 \rangle\), \(\langle \theta^2 \rangle\), and \(\langle z^2 \rangle\) in terms of \(L_p\).

What we obtained in Eq.~(\ref{eq:rFJC_DE}) is the differential equation for the joint probability distribution (positional-orientational) of the rFJC, and it turns out to be somehow different from its counterpart for the WLC. We have the convective derivative in the LHS as in the WLC, but we do not have a second derivative (laplacian) with respect to \(\vec{u}\) in RHS. Since this is not a diffusion-like equation, we cannot make a similar argument to that in the WLC. Thus, it makes it hard to deal with its solution.

Before we go further, we should point out that there are significant similarities and also differences between the rFJC and the WLC. In Section VI, we have obtained several moments for the rFJC. Since the correlation of the tangent vector along the polymer contour decays exponentially, we have the same expression of \(\langle R^2 \rangle\) for both models when expressed in terms of the persistence length \(L_p\). However, the orientation distribution of the last segment and also \(\langle \theta^2 \rangle\) are different because the rFJC behaves "extremely". If the hinge is opened, it allows any orientation uniformly, which is "extreme", unlike the WLC or the KP chain.

From the point of view of the force-extension relation, we expect that the rFJC will have a faster extension than the WLC or KP chain, since closed hinges result in longer rods, which become more likely to be aligned to the direction of the force. Some studies have also introduced reversible bending stiffness in the WLC. Refs.~\onlinecite{benetatos2022ensemble, Razbin2023fluctuatingstiffness} introduce a fluctuating bending stiffness for the rWLC (reversible WLC). It turns out that there is a crossover to the state with the larger value of the bending stiffness when the tensile force is increased. This is similar to the rFJC, where the fraction of closed hinges increases and approaches 1 when the force increases in the thermodynamic limit.

The rod-like limit \(L_p \gg L\) is defined within the continuum limit, implying a non-vanishing orientational correlation between the two ends of the chain. It turns out that in the small force regime, the linear response of the chain (the linear compliance) is proportional to \(\langle z^2 \rangle\), as we expected. We also obtained \(\langle n \rangle\) in the small force regime. The result obtained by this method indicates that if \(\epsilon\) scales as \(\epsilon \sim k_BT\ln(4\pi \alpha N)\), where \(\alpha = L_p/L \gg 1\), then the chain exhibits nonmonotonic behavior in the force-differential compliance and force-occupation number relation.

The other way to calculate the bending elasticity in the thermodynamic limit is the generating function method (necklace model). Using the generating function method, we have obtained the exact force-deflection and the force-occupation number relation for arbitrary values of the applied force, and it turns out that the bending is equal to the stretching. This observation is just a corollary of the generating function method since we have already implied that \(N\) is strictly infinite, {\it i.e.} \(L \gg L_p\), when using the generating function method. Note that the behavior obtained by the generating function method does not give non-monotonic behavior.

It is important to point out that there is no critical value of \(\epsilon\), where non-monotonic behavior suddenly disappears in the thermodynamic limit (\(L_p \ll L\)). The rFJC always exhibits non-monotonic behavior, unless we are strictly at the limit of an infinite chain. For an infinite chain, stretching and bending become strictly identical, and the boundary conditions do not matter. When \(L_p\gg L\), it implies that the microstates with a few "giant" rods dominate in the small force regime, which becomes clear in Eq.~(\ref{eq:meanext_3}). In Eq.~(\ref{eq:meanext_3}), the first term in the second pair of parentheses corresponds to the case where only one hinge is opened so that there are two giant rods. Furthermore, this term gives rise to the increased compliance in the small force regime, Eq.~(\ref{eq:comp_stiff}). This argument also holds for the force-occupation number relation since the argument is based on expanding the partition function.

For a better understanding of our analysis, we summarize the methods that we have employed in the paper before discussing further applications of the model. In Section IV, we have obtained a recursion relation that gives the exact partition function, in principle, for arbitrary \(N\). However, this becomes extremely inefficient when \(N\) increases, since the number of terms grows as \(2^N\). A mean-field approximation introduced in Section V cures this limitation. Ignoring fluctuations of the rod length, the number of terms decreases to \(\mathcal{O}(N^2)\). It also helps to construct a landscape of free energy density, which may be useful for the thermodynamic limit. Despite these advantages, there are some quantitative discrepancies between the mean-field approximation and the exact value, as we have shown in FIG.~\ref{fig:MF_ext} and~\ref{fig:MF_occ}.

The generating function method introduced in Section VII gives an exact description of the thermodynamic limit. The method is based on the assumption that \(N \rightarrow \infty\), keeping other control variables finite (\(f\) and \(\epsilon\) for the rFJC). This becomes exact since we construct a generating function and compute the exact partition function using the method of steepest descent (saddle-point method) for infinitely large \(N\), that is exact in a well-defined thermodynamic limit.

The rFJC has been proposed to deal with a reversible local bending stiffness (zero or infinite). The reversible hinges are opened and closed, controlled by an activation energy \(\epsilon\). If the hinges are closed, the rods linked to the hinge become a longer rod and stiffen. However, for some biopolymers, there may be cooperative phenomena in the local bending stiffness. This can be taken into account by introducing an Ising-type interaction between segments, as in Ref.~\onlinecite{benedito2018isotensional}. The model of Ref.~\onlinecite{benedito2018isotensional} also introduces bistable elasticity, which plays a similar role to the opening-closing of hinges by an activation energy in our model. Thus, it would be interesting to introduce an Ising-type interaction between the segments also for the rFJC, in a future work.

The calculations presented in this paper are all done in the Gibbs ensemble, and it is well known that polymers away from the thermodynamic limit often exhibit statistical ensemble inequivalence (Gibbs versus Helmholtz). Nonetheless, we expect that there will be ensemble equivalence in the thermodynamic limit. This has been investigated for several models of polymers, including the WLC.\cite{winkler2010equivalence, manca2014equivalence,benetatos2022ensemble}

\begin{acknowledgments}
We thank Geunho Noh, Juhyeon Lee, and Soi Ji for their useful comments and discussion. We acknowledge the support of a grant from the National Research Foundation of Korea, NRF-2022R1F1A1070341, funded by the Ministry of Science and ICT, Korea (MSIT).
\newpage

\end{acknowledgments}
 
\appendix

\section{\label{sec:kinetic_appendix}The contribution of the kinetic energy in the partition function}
In the partition function, the relative contribution of the kinetic energy becomes negligible, as we show explicitly in this Appendix. In the spherical coordinate system, the Hamiltonian of the freely rotating rigid rod, {\it i.e.} the kinetic part of the Hamiltonian for polymer rod, reads
\begin{eqnarray}
    H=\frac{1}{2I}\Biggr( p_\phi^2+\frac{p_\theta^2}{\sin^2\theta} \Biggl)
    \label{eq:kinetic_hamiltonian},
\end{eqnarray}
where \(\phi\) is the azimuthal angle, \(\theta\) is the polar angle, and \(I\) is the moment of inertia. Then, the partition function is obtained as
\begin{eqnarray}
    Z_{\text{K.E.}}&=&\frac{1}{\hbar^2} \iint \,d\theta\,d\phi\,dp_\theta\,dp_\phi e^{-\beta H(\theta,\phi,p_\theta,p_\phi)} \nonumber \\
    &=&\frac{2I}{\beta h^2} \qquad (\beta = 1/k_BT).
    \label{eq:kinetic_partit}
\end{eqnarray}
Note that \(I\), the moment of inertia, is proportional to the square of the length of the rod \(j^2\). For the sake of simplicity, we write it as \(Z_{\text{K.E.}} = C j^2\), where \(C\) is a constant.

Multiplying this factor to the partition function for a length \(j\) polymer rod, we have the following.
\begin{eqnarray}
    g(j)&=&C j^2 e^{(j-1)\epsilon}\frac{4\pi \sinh(jf)}{jf}\nonumber \\
    &=& C e^{(j-1)\epsilon + 2\ln j}\frac{4\pi \sinh(jf)}{jf}.
    \label{eq:kinetic_neg}
\end{eqnarray}
We see that the contribution of the kinetic energy becomes negligible compared to that of the bending force and the activation energy. To mention it explicitly, the former scales by polynomial order, and the latter scales exponentially.

\section{\label{sec:ty_appendix}The Rod-like limit using Taylor Expansion of the partition function}

The calculations of Section~\ref{sec:Rod_Lim} are done as follows. The Taylor expansion of the partition function Eq.~(\ref{eq:recur}) with respect to \(f\) yields
\begin{widetext}
    \begin{eqnarray}
        Z(N)&=&(e^\epsilon+4\pi)^N+\frac{f^2}{6}\Biggr[(4\pi)^N N + (4\pi)^{N-1}e^\epsilon(N-1)(N+3) + \dots \nonumber \\
        &+& (4\pi)^2e^{(N-2)\epsilon}\frac{N^2(N+1)(N-2)}{6} + 4\pi e^{(N-1)\epsilon}\frac{N(N+1)(2N+1)}{6}\Biggr] +\mathcal{O}(N^4f^4)
        \label{eq:partit_taylor}.
    \end{eqnarray}
\end{widetext}
The first term in the RHS above is the partition function for the unperturbed chain. Now, to obtain \(\langle z \rangle\) and \(\langle n \rangle\), again, we should expand \(\partial_f Z(N)/Z(N)\) and \(\partial_\epsilon Z(N)/Z(N)\) respectively. Then, for the mean deflection, we obtain
\begin{widetext}
    \begin{eqnarray}
        \langle z \rangle &\approx& \frac{1}{(e^\epsilon+4\pi)^N}\cdot\frac{f}{3}\Biggr[ (4\pi)^N N + (4\pi)^{N-1}e^\epsilon(N-1)(N+3) + \dots \nonumber \\
        &+& (4\pi)^2e^{(N-2)\epsilon}\frac{N^2(N+1)(N-2)}{6} + 4\pi e^{(N-1)\epsilon}\frac{N(N+1)(2N+1)}{6}\Biggr]
        \label{eq:meanext_1}.
    \end{eqnarray}
\end{widetext}

Now, we consider the rod-like limit condition \(\epsilon \approx k_BT \ln(4\pi \alpha N)\). (\(L_p/b\approx e^{\epsilon/k_BT}/4\pi\) in this limit.) Substituting into Eq.~(\ref{eq:meanext_1}), yields
\begin{widetext}
    \begin{eqnarray}
    \langle z \rangle \approx \exp(-\frac{1}{\alpha})\cdot\frac{fb^2}{3k_BT}\Biggr[\,N + \dots + \frac{1}{\alpha^2}\frac{(N+1)(N-1)}{6}+\frac{1}{\alpha}\frac{(N+1)(2N+1)}{6}\Biggr]
    \label{eq:meanext_2}.
    \end{eqnarray}
\end{widetext}

Finally, we arrive at the force-deflection relation at the rod-like limit by replacing \(1/\alpha\) as \(L/L_p\),

\begin{eqnarray}
    \langle z \rangle &\approx& \Biggr[ \frac{L^3}{9 L_p}-\frac{L^4}{18 L_p^2}+ \mathcal{O}\Biggr(\frac{L^5}{L_p^3} \Biggr) \Biggr] \frac{f}{k_BT}\nonumber \\&+& \Biggr[\frac{L^5}{150 L_p} + \mathcal{O}\Biggr(\frac{L^6}{L_p^2} \Biggr) \Biggr] \frac{f^3}{k_B^3T^3}
    \label{eq:meanext_3}.
\end{eqnarray}

If one wants to consider higher order terms of \(f\) for the mean deflection or the mean occupation number, it is also possible by calculating the higher order terms in Eq.~(\ref{eq:partit_taylor}). Here, we have shown the calculations only considering the lea. Note that Eq.~(\ref{eq:meanext_3}) is the result obtained by considering up to the second order.

\section{\label{sec:gf_appendix}The force-deflection and force-occupation number relation yielded by Generating Function method}

In this appendix, we show the exact force-deflection and force-occupation number relation obtained by the generating function method by taking \(\langle z \rangle/L = - \partial_f \ln p^*\) and \(\langle n \rangle = - \partial_\epsilon \ln p^*\). The results are written below. Note that we have restored the units.
\begin{widetext}
    \begin{eqnarray}
        \frac{\langle z \rangle}{L} &=& \frac{\exp(\frac{f b e^{\frac{\epsilon }{k_B T}}}{2 \pi  k_B T}) \left(2 \pi  \exp(\frac{f b \left(e^{\frac{\epsilon }{k_B T}}+4 \pi \right)}{2 \pi  k_B T})-\left(e^{\frac{\epsilon }{k_B T}}+2 \pi \right) \left(e^{\frac{2 f b}{k_B T}}-1\right)\right)-2 \pi }{2 \pi  \left(\exp(\frac{f b e^{\frac{\epsilon }{k_B T}}}{2 \pi  k_B T})-1\right) \left(\exp(\frac{f b \left(e^{\frac{\epsilon }{k_B T}}+4 \pi \right)}{2 \pi  k_B T})-1\right)} \nonumber \\
        \langle n \rangle&=&1-\frac{f b \left(e^{\frac{2 f b}{k_B T}}-1\right) \exp(\frac{f b e^{\frac{\epsilon }{k_B T}}+2 \pi  \epsilon }{2 \pi  k_B T})}{2 \pi  k_B T \left(\exp(\frac{f b e^{\frac{\epsilon }{k_B T}}}{2 \pi  k_B T})-1\right) \left(\exp(\frac{f b \left(e^{\frac{\epsilon }{k_B T}}+4 \pi \right)}{2 \pi  k_B T})-1\right)}
        \label{eq:thm_def_occ}.
    \end{eqnarray}
\end{widetext}

\bibliography{bending_rFJC}
\end{document}